\documentclass[10pt, twocolumn, comsoc]{IEEEtran}

\usepackage{graphicx,epsfig}
\usepackage[noadjust]{cite}
\usepackage{mcite}
\usepackage{amsfonts,helvet}
\usepackage{fancyhdr}
\usepackage{threeparttable}
\usepackage{epsf,epsfig}
\usepackage{amsthm}
\usepackage{amsmath}
\usepackage{siunitx}
\usepackage{amssymb}
\usepackage{dsfont}
\usepackage{subfigure}
\usepackage{color}
\usepackage[linesnumbered,ruled]{algorithm2e}
\usepackage{enumerate}
\usepackage{cancel}
\usepackage{mathptmx}
\usepackage{comment}
\usepackage{siunitx}
\usepackage{xcolor}

\usepackage[normalem]{ulem}

\newtheorem{theorem}{Theorem}

\newtheorem{corollary}{Corollary}

\newtheorem{lemma}{Lemma}

\newtheorem{remark}{Remark}

\SetKwInput{KwInput}{Initialize}
\SetKwInput{KwOutput}{Output}
\SetKwProg{Fn}{Stage 1}{}{}
\SetKwProg{Fn}{Stage 2}{}{}

\usepackage{eucal}

\begin{document}

\title{Multibeam Satellite Communications with Massive MIMO: Asymptotic Performance Analysis and Design Insights}

\author{Seyong~Kim, Jinseok~Choi, Wonjae~Shin, Namyoon~Lee, and Jeonghun~Park

\thanks{
This work was supported by Samsung Research Funding \& Incubation Center of Samsung Electronics under Project Number SRFC-IT2402-06.
S. Kim and J. Park are with the School of Electrical and Electronic Engineering, Yonsei University, Seoul 03722, South Korea (e-mail: {\texttt{sykim@yonsei.ac.kr; jhpark@yonsei.ac.kr}}). J. Choi is with School of Electrical Engineering, Korea Advanced National Institute of Science and Technology, Republic of Korea (e-mail: {\texttt{jinseok@kaist.ac.kr}}). W. Shin is with the School of Electrical Engineering, Korea University, Seoul 02841, South Korea (email: {\texttt{wjshin@korea.ac.kr}}). N. Lee is with Department of Electrical Engineering, POSTECH, Pohang, South Korea (email: {\texttt{nylee@postech.ac.kr}}).}
} 

\maketitle \setcounter{page}{1} 

\begin{abstract} 

Multibeam satellite communication systems are promising to achieve high throughput. 
To achieve high performance without substantial overheads associated with channel state information (CSI) of ground users, we consider a fixed-beam precoding approach, 
where a satellite forms multiple fixed-beams without relying on CSI, then selects a suitable user set for each beam. 
Upon this precoding method, we put forth a satellite equipped with massive multiple-input multiple-output (MIMO), by which inter-beam interference is efficiently mitigated by narrowing the corresponding beam width. 
By modeling the ground users' locations via a Poisson point process, we rigorously analyze the achievable performance of the presented multibeam satellite system. 
In particular, we investigate the asymptotic scaling laws that reveal the interplay between the user density, the number of beams, and the number of antennas. 
Our analysis offers critical design insights for the multibeam satellite with massive MIMO: 
i) If the user density scales proportionally with the number of antennas, the considered precoding can achieve a linear fraction of the optimal rate in the asymptotic regime. ii) A certain additional scaling factor for the user density is needed as the number of beams increases to maintain the asymptotic optimality. 

\end{abstract}


\section{Introduction}

The exponential growth in satellite communications has led to an increasing demand for more efficient spectrum reuse strategies.
As a solution for this challenge, multibeam satellites have been mainly considered \cite{vaszques:wcommmag:16, perez:spmag:19, khammassi:survey:24}, wherein multiple payloads are sent through multiple feeds, so that independent information is simultaneously delivered to each spot beam on the ground. 
Multibeam satellite communications resemble terrestrial multiple-input multiple-output (MIMO) communication systems with spatial multiplexing. 
As well known in the literature, by encoding independent streams and combining them with precoding, 
spatial multiplexing gains are achieved while suppressing inter-stream interference in conventional terrestrial MIMO systems \cite{arapog:survey:11}. 
Similar to this, multibeam satellites achieve significant spatial multiplexing gains by sending an individual frame for each spot beam.

Unfortunately, in multibeam satellite communications, it is infeasible to completely prevent the signal transmitted from a certain beam from radiating to other adjacent beams. 
This results in the inter-beam interference, severely limiting the achievable performance. 
A promising approach to address this problem is applying MIMO precoding techniques at the satellites \cite{perez:spmag:19, arapog:survey:11, khammassi:survey:24}. 
To be specific, by leveraging channel state information (CSI) obtained from ground users, satellite MIMO precoding vectors are designed so as to mitigate the inter-beam interference. 

\subsection{Related Works}
Several prior work has been conducted in the context of MIMO precoding for multibeam satellite communications. 
{\textcolor{black}{In \cite{li:tcom:2023}, a multibeam GEO satellite equipped with MIMO was considered, where a sum-rate maximization method was proposed under quality of service (QoS) constraints. To this end, users are selected in each time slot, and the corresponding precoding matrix is computed based on channel feedback from the selected user group. 
In \cite{wang:tvt:2021}, multiple beams are exploited in GEO satellite communications to serve multiple users simultaneously. Building on this setup, a robust beamforming algorithm was developed to enhance resource efficiency while satisfying QoS and total power constraints.}}
\textcolor{black}{In \cite{chris:twc:15}, an iterative algorithm was developed for a multigroup multicast precoder in DVB-S2X, which maximizes the sum rate while mitigating interference. 
In \cite{vazquez:jsac:18}, the signal-to-interference-plus-noise ratio (SINR) was characterized by incorporating impacts of the outdated CSI, and a precoding method based on minimum mean square error (MMSE) to mitigate multibeam interference while accounting for noise. In \cite{schwarz:tbc:19}, a fixed MIMO satellite system was studied under line-of-sight (LoS) satellite channels considering inter-stream interference.  
In multiple gateways with multibeam satellite communications, \cite{joroughi:twc:16} considered full frequency reuse with interference mitigation.}
A comprehensive survey on the state-of-art precoding techniques for multibeam satellites is found in \cite{khammassi:survey:24}.

{\color{black}{Despite the significant gains offered by the aforementioned MIMO precoding approaches, 
their practical implementation in multibeam satellite communications faces several key challenges.
First, accurate CSI is hard to obtain. In multibeam satellite communications, obtaining precise CSI is particularly challenging due to long propagation distances and significant signal attenuation. It is well known that MIMO spatial multiplexing gains are eroded by the CSI acquisition error \cite{park:twc:16, jindal:tvt:06, ahmad:tvt:21}. 
Second, the high computational complexity of precoding optimization is critical challenges in satellite communications. For instance, \cite{chatzinotas:asilomar:11} employed regularized zero-forcing (RZF), which involves matrix inversion operations whose computational complexity increases cubically with the size of the matrix.
\cite{lei:tvt:2024} employed CVX tools to solve an iterative algorithm based on semi-definite programming (SDP) for GEO satellite communications; however, this approach incurs substantial computational overhead, making it less suitable for practice.
Third, the design of MIMO precoders typically requires joint gateway processing, where the CSI of all users must be aggregated and processed centrally.
This joint operation necessitates the precoding computation to be performed either within a single gateway or through on-board processing on the satellites. 
The joint gateway processing can not only introduce additional signal processing delay, but also require more bandwidth to share the CSI between gateways \cite{na:jcn:2023}. In light of these challenges, 3GPP NTN has adopted fixed-beam precoding in GEO satellite communications \cite{3gpp:NTN:2023}.}}

{\textcolor{black}{To mitigate the inter-beam interference while circumventing the above obstacles, an attractive approach is to employ orthogonal fixed-beams predefined at the satellite, combined with a careful selection of ground users.
The fixed-beam precoding approach is closely related to a concept of random beamforming \cite{sharif:tit:05, gilwon:twc:2016} in the multibeam MIMO literature, which aims to obtain multi-user diversity. 
A key principle of multi-user diversity is that, although inter-beam interference is unavoidable with predefined beams, it is possible to achieve near-optimal performance by selecting a set of users well-suited for those beams, provided that the number of users is sufficiently large.}}
This approach has several different names, but we refer to this as a fixed-beam precoding approach to avoid confusion. 
\textcolor{black}{\cite{wu:mag:2024} focused on large-scale MIMO, incorporating key challenges such as limited satellite payload capacity, line-of-sight (LoS) channel conditions, and channel aging.}
In \cite{zorba:space:08}, an improved multibeam opportunistic precoding method was proposed, that only requires partial CSI. 
In \cite{koyoungcahi:twc:2022}, a satellite employed a fixed precoder and selected ground users based on signal-to-noise ratio (SNR), modeling their spatial locations with a Poisson point process (PPP) to analyze coverage probability using stochastic geometry tools.

In the fixed-beam precoding, the beam width primarily determines the inter-beam interference levels. Specifically, narrower beams reduce interference, while wider beams increase it between adjacent beams. 
From this point of view, a massive MIMO technology \cite{larsson:commmag:14, rusek:spmag:13} is particularly beneficial when used with the fixed-beam precoding approach in multibeam satellite systems. The large-scale antenna array enables the creation of extremely narrow beams, which is useful to mitigate inter-beam interference \cite{ngo:eusipco:2014}. 
Motivated by this, \cite{angeletti:access:20} proposed to adopt massive MIMO into a multibeam satellite system, and presented a switchable fixed multibeam strategy. 
Despite the promising potential of massive MIMO in multibeam satellite communications, there remains a paucity of rigorous performance analyses that offer comprehensive analytical insights into the system's behavior in the massive MIMO regime.  
Notably, the interplay between the array size and the spatial density of ground users has yet to be fully explored.
Such an in-depth analysis is crucial, as it would provide invaluable design guidelines for deploying massive MIMO technology in multibeam satellite communication systems. 
This paper aims to fill this knowledge gap by conducting a comprehensive performance analysis of multibeam satellite communications equipped with massive MIMO.




\subsection{Motivations and Contributions}

In this paper, we consider a GEO downlink (i.e., forward link) multibeam satellite communication system equipped with massive MIMO. In the considered setup, we assume that a GEO satellite serves multiple ground users, by sending independent information for each spot beam. 

\textcolor{black}{Specifically, we consider the fixed-beam precoding approach
for multibeam satellite communication systems. This contrasts to conventional MIMO precoding, which depends on instantaneous CSI and incurs significant computational and centralized processing overhead \cite{li:tcom:2023,wang:tvt:2021,chris:twc:15,vazquez:jsac:18,joroughi:twc:16,chatzinotas:asilomar:11,lei:tvt:2024}.
Distinguished from this, in our approach, the GEO satellite exploits a large-scale antenna array to form multiple fixed-beams directed toward predetermined spatial locations. Notably, this precoding process does not require instantaneous CSI, resulting in low computational complexity.} 
Incorporating the multiple fixed-beams, we select a proper set of ground users, one per each spot beam. In this user selection process, we only exploit the spatial location of the ground users, which is corresponding to the long-term CSI of the ground users. 


    {\textcolor{black}{Upon this setup, our main contribution is to derive the scaling laws of the ergodic rate in the asymptotic regime, expressed in a concise closed-form. 
    This captures the system’s behavior according to the key parameters (e.g. user density and the number of antennas and beams), providing valuable guidance for the design criteria of overall networks. Additionally, we include the ergodic rate analyses based on \cite{hamdi:tcom:10} for the completeness of the paper. This can provide readers with a broader perspective and additional insights into our analysis.}}
    
    From these analyses, we draw key design insights for multibeam satellite systems with massive MIMO:
    (i) In a single beam scenario, if the user density scales proportionally with the number of antennas, the fixed-beam approach with location-based user selection achieves a linear fraction of the optimal rate, even without user's CSI. If the user density and the number of antennas scale equally, the fixed-beam method achieves the same asymptotic rate scaling as the optimal approach.
    (ii) In a multibeam scenario, we quantify the probability of maintaining interference below a given threshold and provide necessary conditions for achieving this based on beam spacing and user density.
    (iii) In a multibeam scenario, to further mitigate inter-beam interference, the user density requires an additional scaling factor compared to the single beam scenario. 
    In this regard, we present a clear relationship between the number of antennas, the number of beams, and the user density. 
    These findings provide useful guidelines for designing fixed-beam multibeam satellite systems with massive MIMO.

{\textcolor{black}{Our work is relevant to several prior studies examining the performance of satellite communication systems \cite{koyoungcahi:twc:2022, park:twc:2023, kim:twc:24, jung:tcom:22, lim:twc:2024, talgat:taes:24}. 
For clarification, we explain our distinguishable contributions compared to the existing work.
In \cite{jung:tcom:22, lim:twc:2024, talgat:taes:24, park:twc:2023, kim:twc:24}, a low Earth orbit (LEO) satellite network was investigated, assuming the spatial distribution modeled by a PPP.
In \cite{jung:tcom:22}, multiple LEO satellites were distributed over a sphere according to a homogeneous binomial point process (BPP), and by applying the Poisson limit theorem, the satellite distribution was asymptotically approximated to a PPP, yielding outage probability results.
In \cite{lim:twc:2024}, the outage probability was derived by capturing the characteristic of the interference from terrestrial networks to a satellite receiver. 
In \cite{talgat:taes:24}, the coverage probability under a Shadowed-Rician fading channel was analyzed, assuming a Poisson cluster process for the user distribution.
However, the analyses in \cite{jung:tcom:22, lim:twc:2024, talgat:taes:24, park:twc:2023, kim:twc:24} did not fully explore the impact of the number of antennas in satellite communication environments. In contrast, our work addresses this gap by rigorously analyzing the beam gains as a function of both the number of antennas and user density, and deriving the ergodic rate scaling laws in a concise closed-form expression.}} In addition, \cite{koyoungcahi:twc:2022} shares a similar scope to our work in that the fixed-beam precoding is applied with user selection. 
However, our unique contribution lies in conducting a rigorous scaling law analysis, which provides insights into system behavior as the number of antennas and the user density increase—which was not addressed in \cite{koyoungcahi:twc:2022}. While a scaling law analysis was performed in \cite{gilwon:twc:2016}, it focused only on a 2D terrestrial network with a uniform linear array (ULA), which cannot be extended to the 3D network scenarios relevant to multibeam satellite communications. To the best of our knowledge, no prior work has performed an asymptotic performance analysis considering a 3D network scenario applicable to GEO satellite communications.


\textcolor{black}{\textbf{Notations}: The following notations are used throughout this paper. Vectors are represented by bold lowercase letters. The Kronecker product of two vectors $\mathbf{a}$ and $\mathbf{b}$ is denoted by $\mathbf{a} \otimes \mathbf{b}$. The modulus and Frobenius norm are denoted by $|\cdot|$ and $\| \cdot \|$, respectively. $\mathbb{E}[\cdot]$ denotes the expected value of a random variable and $\mathbb{P}[\cdot]$ indicates the probability measure. The probability density function (PDF) of a random variable $X$ is denoted by $f_X(x)$. For a complex value $g$ that follows a Shadowed-Rician (SR) distribution, we write $g \sim \mathcal{SR}(\Omega, b_0, m)$, where $\Omega$ is the average power of the LoS component, $2b_0$ is the average power of the scattered components, and $m$ is the Nakagami parameter. $\mathbb{Z}$ is the set of integers.
}

\section{System Model}
\begin{figure}[t]
    \centerline{\resizebox{0.8\columnwidth}{!}{\includegraphics{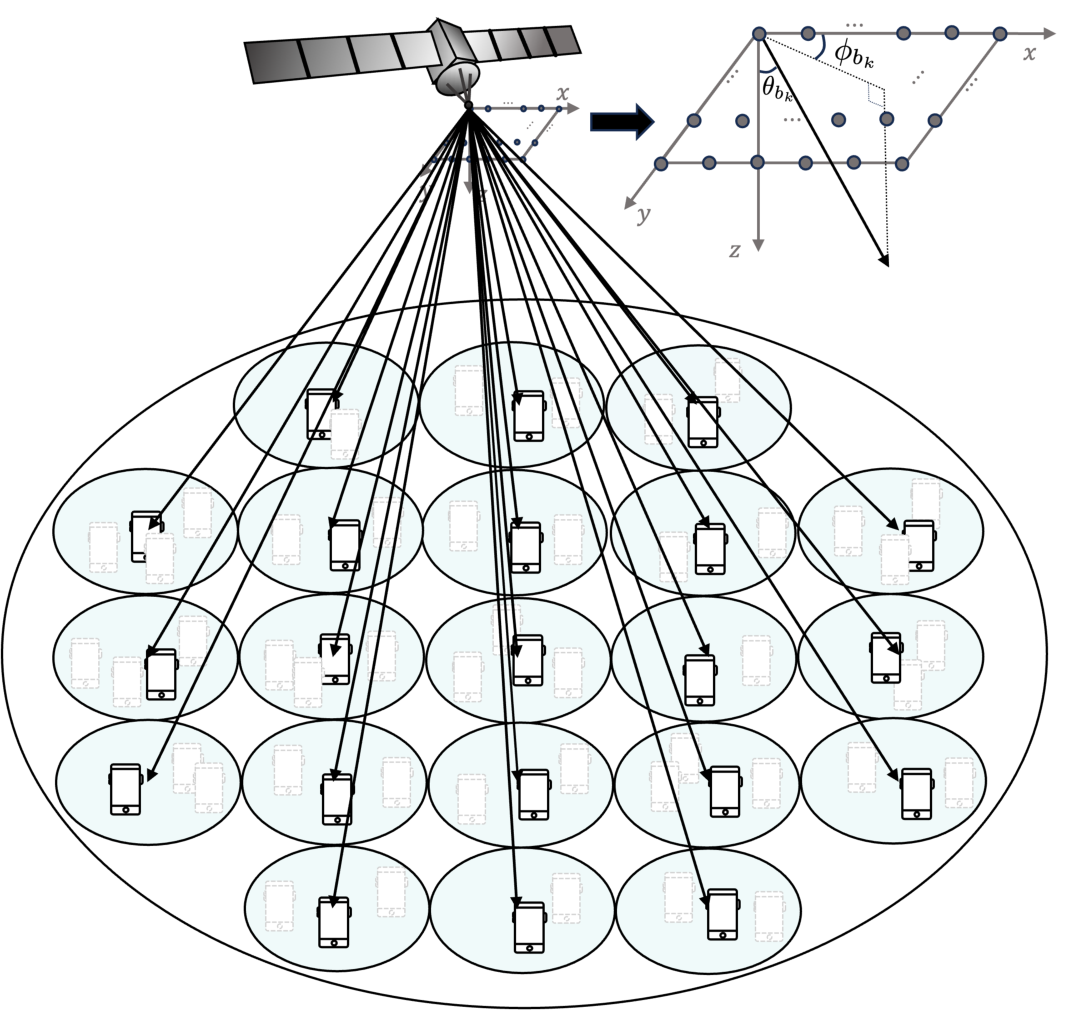}}} 
        \caption{Illustration of the downlink multibeam satellite communication.}
        \label{fig:system model}
\end{figure}

We consider a downlink GEO multibeam satellite communication system, where the satellite is equipped with uniform planar arrays (UPAs) that consist of $M_{\text x}$ number of antennas arranged along the $\text x$-axis and $M_{\text y}$ number of antennas along the {\text y}-axis. We assume $M_{\text{x}} = M_{\text y} = M$, thereby the total number of antennas is $M^2$. 
For explanations regarding the considered array model, we refer to Remark \ref{remark:array model}. 
The satellite forms $K\ge 1$ number of beams, wherein $K$ number of single-antenna users are served by each beam, i.e., a single user per beam.
We also assume that each beam shares the same time-frequency resource, i.e., full frequency reuse. 
The more detailed setup is as follows.

\subsection{Network Model}
We assume that the ground users are spatially distributed by a homogeneous PPP denoted by $\Phi = \{{\bf{d}}_i \in \mathbb{R}^2, 1 \le i \le N\}$ with a uniform intensity $\lambda$. 
Denoting that the whole coverage region of the considered satellite as a disk with radius $R_{\text{cov}}$, $N$ follows the Poisson distribution with mean $\lambda \pi R_{\text{cov}}^2$. 
For convenience, we let the whole coverage region be $\CMcal{A}$. 
Additionally, we assume that each beam covers a designated region denoted as $\CMcal{A}_k$, and the coverage region of the $k$-th beam is a disk with radius $R_k$ for $1 \le k \le K$. 
Accordingly, the average number of users included in the $k$-th beam's coverage region is $\lambda \pi R_k^2$. 
Our network model is illustrated in Fig.~\ref{fig:system model}.


\begin{table}[t] 
    \centering
    \caption{SR Fading Parameters}
    \label{table:SR}
    \begin{tabular}{|>{\centering\arraybackslash}m{3.2cm}|>{\centering\arraybackslash}m{1.4cm}|>{\centering\arraybackslash}m{0.7cm}|>{\centering\arraybackslash}m{0.7cm}|}
    \hline
    \textbf{Shadowing Scenario} & $\Omega$ & $b_0$ & $m$ \\
    \hline
    \hline
    Frequent heavy shadowing & $8.97 \times 10^{-4}$ & 0.063 & 0.739 \\
    \hline
    Infrequent light shadowing & 1.29 & 0.158 & 19.4 \\
    \hline
    Average shadowing & 0.835 & 0.126 & 10.1 \\
    \hline
    \end{tabular}
\end{table}

\subsection{Channel Model}

We describe the large-scale fading, the small-scale fading, and the array steering vector as follows. 

    {\textbf{Large-scale fading}}: 
    For user $i$, the large-scale path-loss gain is given by
    \begin{align}\label{eq:LSF}
        L_{i} = \left(\frac{c_0}{4\pi f_c d_i} \right)^2,
    \end{align}
    where $c_0$ and $f_c$ denote the speed of light and the carrier frequency. Also, $d_i$ is the distance from the satellite to user $i$.

    {\textbf{Small-scale fading}}: We let the small-scale fading drawn from the SR distribution. 
    Note that the SR distribution is known to suitably reflect the satellite propagation environments, as shown in \cite{talgat:taes:24,koyoungcahi:twc:2022,Sellathurai:asilomar:2016}. 
    For $g \sim \CMcal{SR}(\Omega, b_0, m) $, the fading power is denoted as $X = |g|^2$ whose PDF is given by
\begin{align} \label{eq:SSF}
    f_{X}(x) =  \left(\frac{2b_0 m}{2b_0m + \Omega} \right)^2 \frac{e^{-\frac{x}{2b_0}}}{2b_0}  {}_1F_1\left(m,1,\frac{\Omega x}{2b_0(2b_0m+\Omega)}\right),
\end{align}
as presented in \cite{abdi:twc:2003}. 
In \eqref{eq:SSF}, ${}_1F_1$ is a confluent hypergeometric function of the first kind.
{\textcolor{black}{To consider the impact of system parameters, we simulate three different SR fading scenarios specified in Table \ref{table:SR}, as referred to in \cite{koyoungcahi:twc:2022,talgat:taes:24}.}}

{\textbf{Array steering vector}}: Considering the UPA, we define the array steering vector as
\begin{align}
\mathbf{v}_i \triangleq \mathbf{v} (\vartheta^x_i) \otimes \mathbf{v} (\vartheta^y_i), \nonumber
\end{align}
where $\otimes$ denotes the Kronecker product and $\mathbf{v}(\cdot)$ is given by
\begin{align}
    \mathbf{v}(x) = \frac{1}{\sqrt{M}} \left[1, e^{-j\pi \frac{2d}{\lambda_0}x}, \cdots, e^{-j\pi(M-1)\frac{2d}{\lambda_0}x} \right]^{\sf T}, \nonumber
\end{align}
and
\begin{align}\label{eq:vartheta}
    \vartheta_{i}^x &=  \sin{\theta_{i}} \cos{\phi_{i}} \notag \\
    \vartheta_{i}^y &=  \sin{\theta_{i}} \sin{\phi_{i}}.
\end{align}
Here, $d$ and $\lambda_0$ are the inter-antenna spacing and the carrier wavelength, respectively. We adopt the half-wavelength antenna spacing, i.e., $d= \frac{\lambda_0}{2}$.
Given that the GEO satellite is geostationarily positioned at the nadir of the center of the coverage region, $\theta_{i}$ and $\phi_{i}$ are the elevation angle and the azimuth angle of the user $i$ respectively, as depicted in Fig.~\ref{fig:system model}

Combining the large-scale fading, the small-scale fading, and the array steering vector, the propagation channel for user $i$ is modeled as
\begin{align}\label{eq:channel}
    \mathbf{h}_i = \sqrt{L_i} g_i M \mathbf{v}_{i} \in \mathbb{C}^{M^2 \times 1},
\end{align}
where $L_i$ is the large-scale fading defined in \eqref{eq:LSF} and $g_i$ is the small-scale fading defined in \eqref{eq:SSF}. 
\textcolor{black}{We assume the LoS channel as in \eqref{eq:channel}. We also note that this assumption has been adopted in prior work \cite{zheng:twc:2012,angeletti:access:20}.
This assumption is justified by the fact that if a propagation distance significantly exceeds a region of reflection reaching a user, the reflected path length becomes negligible, effectively resulting in LoS channel. Additionally, we assume that delay effects are compensated by performing frequency and time synchronization at each user to employ downlink wideband transmission \cite{li:tcom:22}.}

{\textcolor{black}{
\begin{remark} [Rain attenuation] \normalfont 
Rain attenuation is one of the factors affecting satellite communication performance. 
It typically exhibits spatial correlation over tens of kilometers and changes very slowly \cite{zheng:twc:2012}. 
Since our fixed-beam precoding approach selects the user for beam $k$ within the coverage region $\CMcal{A}_k$, 
it is feasible to assume that the candidate users in $\CMcal{A}_k$ experience identical rain attenuation. 
For this reason, rain attenuation remains constant and does not influence the asymptotic scaling analysis. 
This assumption aligns with the approach taken in \cite{koyoungcahi:twc:2022}.
\end{remark}
}}

\subsection{Precoding Model}
Next, we explain the precoding model. 
By incorporating the beam pattern of the UPA \cite{van:book:2002}, 
we divide the coverage area into a uniform grid and place the beam centers at each grid point. 
For example, the center point of the $k$-th beam is configured as
\begin{align}
    \vartheta_{k}^x = \sin \theta_k \cos \phi_k = \frac{2n}{M^\ell} \nonumber \\
    \vartheta_{k}^y = \sin \theta_k \sin \phi_k = \frac{2m}{M^\ell}, \label{eq:beam position}
\end{align}
where $\{n,m\} \in \mathbb{Z}$ and $\theta_k$ and $\phi_k$ are elevation angle and azimuth angle of the $k$-th beam, respectively. 
$\ell$ is a parameter that adjusts the beam spacing. By increasing $\ell$, the beam spacing becomes narrower, and by decreasing $\ell$, the beam spacing becomes wider. Specifically, when $\ell=1$, each beam is positioned at the first null point of the adjacent beam's pattern, where each beam covers the region of a disk with radius $R_k = \frac{H}{\sqrt{M^2-1}}$ where $H$ is the altitude of the GEO satellite.
A detailed analysis of beam spacing and interference levels will be conducted in Section \ref{sec:multibeam}. 
According to this beam construction, the precoding vector for the $k$-th beam, denoted as ${\bf{f}}_k$, is formed as 
\begin{align}\label{eq:precoding vector}
    \mathbf{f}_{k} = \mathbf{v} (\vartheta^x_{k}) \otimes \mathbf{v} (\vartheta^y_{k}).
\end{align}
We clarify that the precoding vectors do not change depending on the CSI of the ground users. 
Without loss of generality, we denote the coverage region corresponding to the $k$-th beam as $\CMcal{A}_k$, so that $\bigcup_{k =1}^{K} \CMcal{A}_k \subseteq \CMcal{A}$, where $|\CMcal{A}_k| = 2\pi R_k^2$. 
Notice that we consider digital precoding, so that multiple payloads are precoded and sent simultaneously.




Subsequently, we describe the user selection. In order to select a user for $k$-th beam, we first extract the users located within $\CMcal{A}_k$ and form a candidate set $\Phi_k = \{{\bf{d}}_i \in \CMcal{A}_k\}$. 
In $\Phi_k$, we select the user whose distance to the corresponding beam center is minimum, i.e., 
\begin{align}
    k^* = {\arg \min}_{{\bf{d}}_i \in \Phi_k} \|{\bf{b}}_k - {\bf{d}}_i \|^2,  \label{eq:user_sel}
\end{align}
where ${\bf{b}}_k$ is the spatial location of the $k$-th beam on the ground in \eqref{eq:beam position}. 
After selecting a user per beam, the satellite sends the messages through the predefined precoding vectors ${\bf{f}}_k$. Since we only select the users by exploiting the spatial locations, no instantaneous CSI feedback is required in this stage.
\footnote{
The user selection process may favor only users located in specific spatial regions, especially those located close to the corresponding beam center. To address this, it is possible to design multiple precoder sets, each with beam centers directed towards different spatial locations. Then these sets are used alternatively across time-frequency resources, enabling the satellite to provide ubiquitous coverage. 
This approach is particularly suitable for the considered UPA, as it is very flexible in forming diverse beam patterns. In contrast, a parabolic reflector array requires physically adjusting the reflector, which hinders the formation of various beams.} 
{\textcolor{black}{Since we have fixed precoder directions as configured in \eqref{eq:beam position}, the selected user's location as in \eqref{eq:user_sel} becomes a critical factor affecting the rate performance. For clarity, the location of the user refers to the distance from the center of the beam. The distance from the beam center impacts the achievable beam gain, which is modeled by the $\text{Fej\'{e}r kernel}$.
We further elaborate on this in the following remark.}}







\begin{figure}[t]
    \centerline{\resizebox{1\columnwidth}{!}{\includegraphics{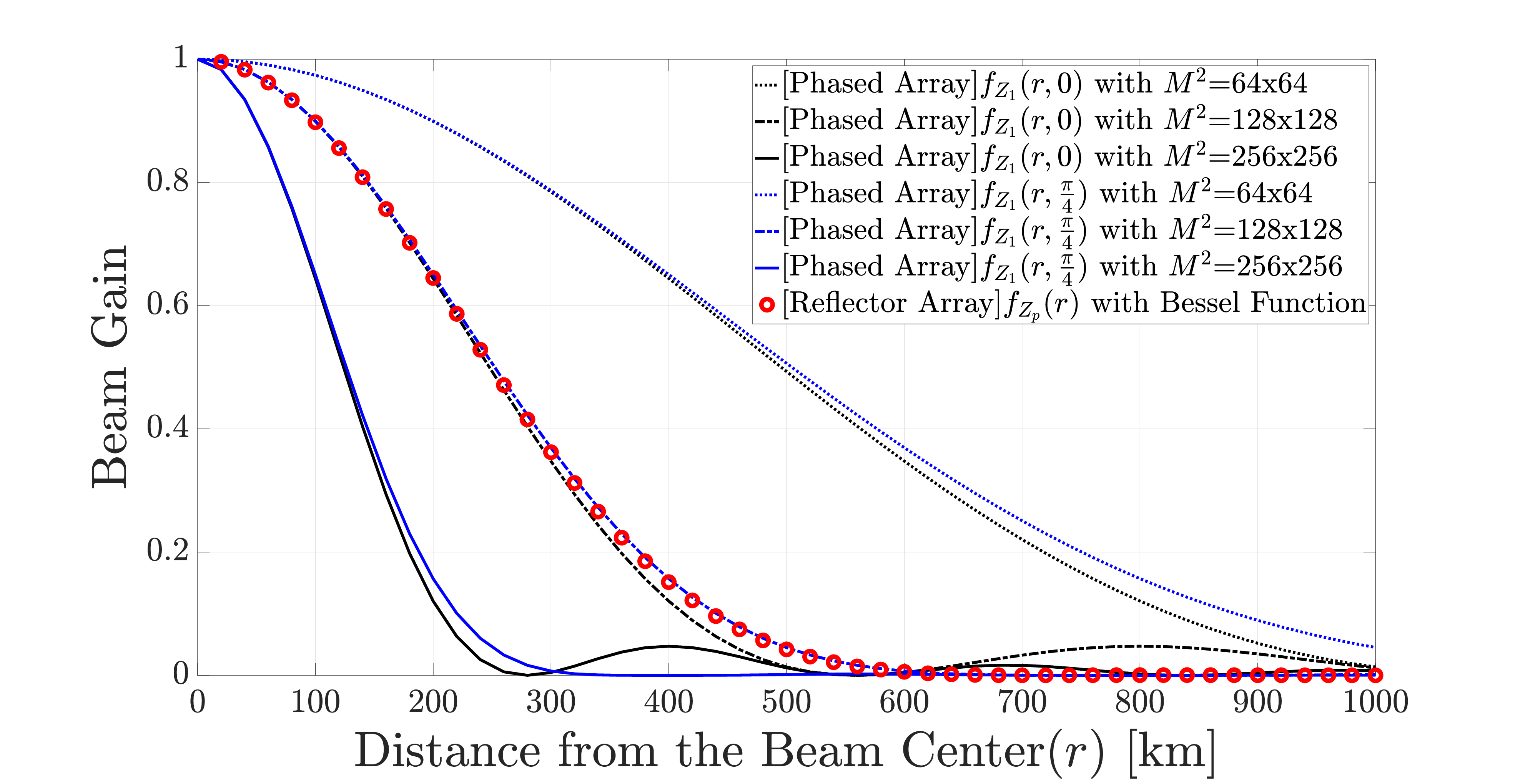}}} 
        \caption{Beam gain comparisons between the parabolic reflector array versus the phased array.}
        \label{fig:beam gain}
\end{figure}

\begin{remark} [Satellite array model] \normalfont \label{remark:array model}


In this remark, we compare the parabolic reflector array commonly considered in the previous studies on GEO multibeam satellites, with the phased array, which is the primary focus of this paper.
The parabolic reflector array is one of the most classic types of directive antenna. It uses a parabolic-shaped reflector to focus the propagated signals, achieving high beam gain with low complexity and low power consumption. 
Due to this benefit, the parabolic reflector array has been commonly used in multibeam satellite systems \cite{koyoungcahi:twc:2022, chatzinotas:asilomar:11,zheng:twc:2012,zorba:space:08}.
Nonetheless, since its beam steering should rely on physically moving the reflector, the parabolic reflector array has limited flexibility in generating and adjusting multiple beams. For instance, considering a single feed per beam case, $K$ number of reflectors are needed to make $K$ spot beams \cite{schneider:ceas:11}.  

In contrast, phased arrays are composed of a large number of small discrete antenna elements arranged in a certain grid, where each element has its own feed and they are controlled electronically.
Because of this feature, beam steering in the phased array is done by electronically adjusting the phase and amplitude of the signals of each antenna element, allowing rapid and precise beam steering without physically moving the aperture. 
The phased array was not popular for satellite communications due to its high complexity and cost. However, with recent advances in phased array hardware and their powerful beam steering capabilities, phased arrays are increasingly considered a viable and beneficial option. This applies not only to LEO satellite communications
\cite{kim:twc:24, you:jsac:20}, but also to GEO satellite communications, as demonstrated in several studies \cite{tian:aces:17, rao:aeromag:13, harris:milcom:00, yu:aeromag:23}. 
This justifies our consideration.  



\textcolor{black}{To make it more understandable, we compare the beam gain functions between the parabolic reflector array and the phased array. In the reflector array, we assume that the tapered-aperture feed reflector is used \cite{lei:tvt:2024,li:tcom:2023,wang:tvt:2021, koyoungcahi:twc:2022,zheng:twc:2012,zorba:space:08}.}
\textcolor{black}{A comprehensive survey on multiple-antenna techniques for satellite systems was presented in \cite{bakhsh:surv:2024}.}
With the nadir-pointing beam, we denote the associated user $1$ whose the distance from beam center to the user is $r_1$ and the azimuth angle is $\phi_1$. Then, the beam gain $ Z_p$ of the parabolic reflector array is approximated as \cite{koyoungcahi:twc:2022, chatzinotas:asilomar:11,zheng:twc:2012,zorba:space:08}
\begin{align}\label{eq:bessel}
    f_{Z_p}(r) = \left|\frac{J_1(u)}{2u} + 36\frac{J_3(u)}{u^3} \right|^2,
\end{align}
where  $J_1$ and $J_3$ are the first-kind Bessel function of order 1 and 3, respectively. 
In addition, $u=2.07123 \sin(\theta_{1})/\sin(\theta_{\text{3dB}})$, where $\sin\theta_{1}=r_1/\sqrt{r_1^2 + H^2}$ and $H$ is the altitude of satellite. $\theta_{{\text{3dB}}}$ is a constant angle associated with the corresponding beam's 3dB angle. 
Under the same assumption, the beam gain $Z_1$ of the phased array is 
\begin{align} 
    f_{Z_1}(r_1,\phi_1) 
    &= \left| \left[ \mathbf{v} (\vartheta_1^x) \otimes \mathbf{v} (\vartheta_1^y) \right]^{\sf H} \left[ \mathbf{v} (0) 
    \otimes \mathbf{v} (0) \right] \right|^2, \nonumber \\
    & \mathop{=}^{\text{(a)}} \frac{1}{M^4}\left| \frac{ \sin\left(\frac{\pi M r_1 \cos \phi_1}{2 \sqrt{ r_1^2 + H^2}}  \right) \sin\left(\frac{\pi M r_1 \sin \phi_1}{2 \sqrt{ r_1^2 + H^2}}  \right)}{ \sin\left(\frac{\pi r_1 \cos \phi_1}{2 \sqrt{ r_1^2 + H^2}}  \right) \sin\left(\frac{\pi r_1 \sin \phi_1}{2 \sqrt{ r_1^2 + H^2}}  \right) }  \right|^2, \label{eq:beam_gain}
\end{align}
where (a) follows that the inner product of two array response vector is represented as $\text{Fej\'{e}r kernel}$ with $| \mathbf{v} (\vartheta_i)^{\sf H}  \mathbf{v} (\vartheta_j) | = F_M(\vartheta_i - \vartheta_j) = \frac{1}{M} \left| \frac{ \sin \frac{\pi M}{2}(\vartheta_i - \vartheta_j) }{\sin \frac{\pi}{2}(\vartheta_i - \vartheta_j)} \right|$ in \cite{gilwon:twc:2016} and $\sin \theta_1 = r_1/\sqrt{r_1^2 +H^2}$. 
The beam gain, which quantifies the alignment between the precoding vector and the user channel, ranges from $0$ to $1$. Fig.~\ref{fig:beam gain} compares the beam gain patterns of a phased array and a reflector array at azimuth angles $\phi=0$ and $\phi=\pi/4$.
In \eqref{eq:beam_gain}, the phased array is represented by $\text{Fej\'{e}r kernel}$, while the reflector array is expressed using the Bessel function, as shown in \eqref{eq:bessel}. As shown in Fig.~\ref{fig:beam gain}, the phased array with $M^2 = 64 \times 64$ achieves a beam gain pattern similar to that of the parabolic reflector array. Moreover, the phased array shows a reduction in beam width as the number of antennas increases. Consequently, to fully leverage the fixed-beam approach, it is necessary to have users positioned closer to the beam center as the number of antennas increases, which can be interpreted as requiring a higher user density.
This relationship aligns directly with our asymptotic findings, highlighting the connection between user density and ergodic rate with fixed-beam precoding as presented in the remainder of this paper.
\end{remark}

\section{A Single Beam Case}
{\textcolor{black}{In this section, we first focus on a single fixed-beam scenario, i.e., $K=1$. 
We extend this setup to a multiple beam case in the next section.}}


\subsection{Achievable Rate Analysis}
We consider that the satellite operates a single nadir-pointing fixed-beam and selects a user whose distance is closest from the beam center.
{\color{black}{Selecting the user closest to the beam center has two key purposes: first, it maximizes the beam gain, thereby enhancing the achievable SNR; second, it enables the accurate spatial distribution required for the scaling law analysis.}}
Without loss of generality, the selected user is assigned the index $1$. The reason we analyze the nadir-pointing fixed-beam and the corresponding user is that it causes the largest variation in beam gain for the same distance $r$. Then, the received SNR for user $1$ is given by 
\begin{align}\label{eq:SNR1}
    \text{SNR}_1 
    &= P G_{\text{Tx}} G_{\text{Rx}} L_1 |g_1|^2 M^2 f_{Z_1}(r_1, \phi_1), 
\end{align}
where $G_{\text{Tx}}$ and $G_{\text{Rx}}$ denote the transmit and receiver antenna gains. 
Additionally, $P = \frac{ P_0}{\kappa T B}$ where $P_0$ is the transmit power of the satellite, $\kappa$ is Boltzmann constant, $T$ is temperature in Kelvin and $B$ is bandwidth, respectively. 
We also note that the fading power $|g_1|^2$ is drawn from the PDF \eqref{eq:SSF}. 
$f_{Z_1}(r_1, \phi_1)$ is a beam gain function defined as \eqref{eq:beam_gain} with the nadir-pointing beam, i.e., $\theta_1 = \phi_1 = 0$. 
By leveraging this, we derive the achievable rate of the user $1$ in the following corollary. 
\begin{figure*}
    \begin{align}\label{eq:raw R1}
        \mathcal{R}_1 = \int_0^{R_1} \int_0^{2\pi} \int_0^\infty \frac{1}{\tau} \left( 1 - \frac{(2b_0m)^m \left(1+2 \tau b_0  P G_{\text{Tx}} G_{\text{Rx}} L_1  M^2 f_Z(r, \phi) \right)^{m-1}}{ \left[(2b_0m + \Omega)\left(1+2 \tau b_0  P G_{\text{Tx}} G_{\text{Rx}} L_1 M^2 f_Z(r, \phi) \right)-\Omega \right]^m } \right) 
        \lambda r e^{-\pi \lambda r^2} e^{-\tau} d\tau d\phi dr 
    \end{align}
\noindent\hrulefill 
\end{figure*}
{\textcolor{black}{\begin{corollary}\label{coro: single achievable rate}
In a single beam case, we define the achievable ergodic rate as 
\begin{align}
    \mathcal{R}_1 &= \mathbb{E}\left[ \log\left(1+ P G_{\text{Tx}} G_{\text{Rx}} L_1 |g_1|^2 M^2 f_{Z_1}(r_1, \phi_1) \right)  \right]
\end{align}
where the expectation is regarding the randomness associated with the fading power and the spatial locations of the ground users. Then $\mathcal{R}_1$ is obtained as in \eqref{eq:raw R1}. 
\end{corollary}
\begin{proof}  
    See Appendix \ref{appendix: single achievable rate}
\end{proof}}}

\begin{table}[t] 
\centering
\caption{System Parameters}
\label{table:parameter}
\begin{tabular}{|c|c|}
\hline
\textbf{Parameter} & \textbf{Value} \\
\hline
\hline
Satellite height & $H = 35786$ km \\
\hline
Link frequency band & $f_c = 20$ GHz (Ka) \\
\hline
Beam bandwidth & $B = 500$ MHz \\
\hline
Noise temperature & $T = 517$ K \\
\hline
Boltzmann constant & $\kappa = 1.3807 \times 10^{-23}$ \\
\hline
User antenna gain & 41.7 dBi \\
\hline
Satellite antenna gain & 52 dBi \\
\hline

\end{tabular}
\end{table}

\begin{figure}[t]
    \centerline{\resizebox{1\columnwidth}{!}{\includegraphics{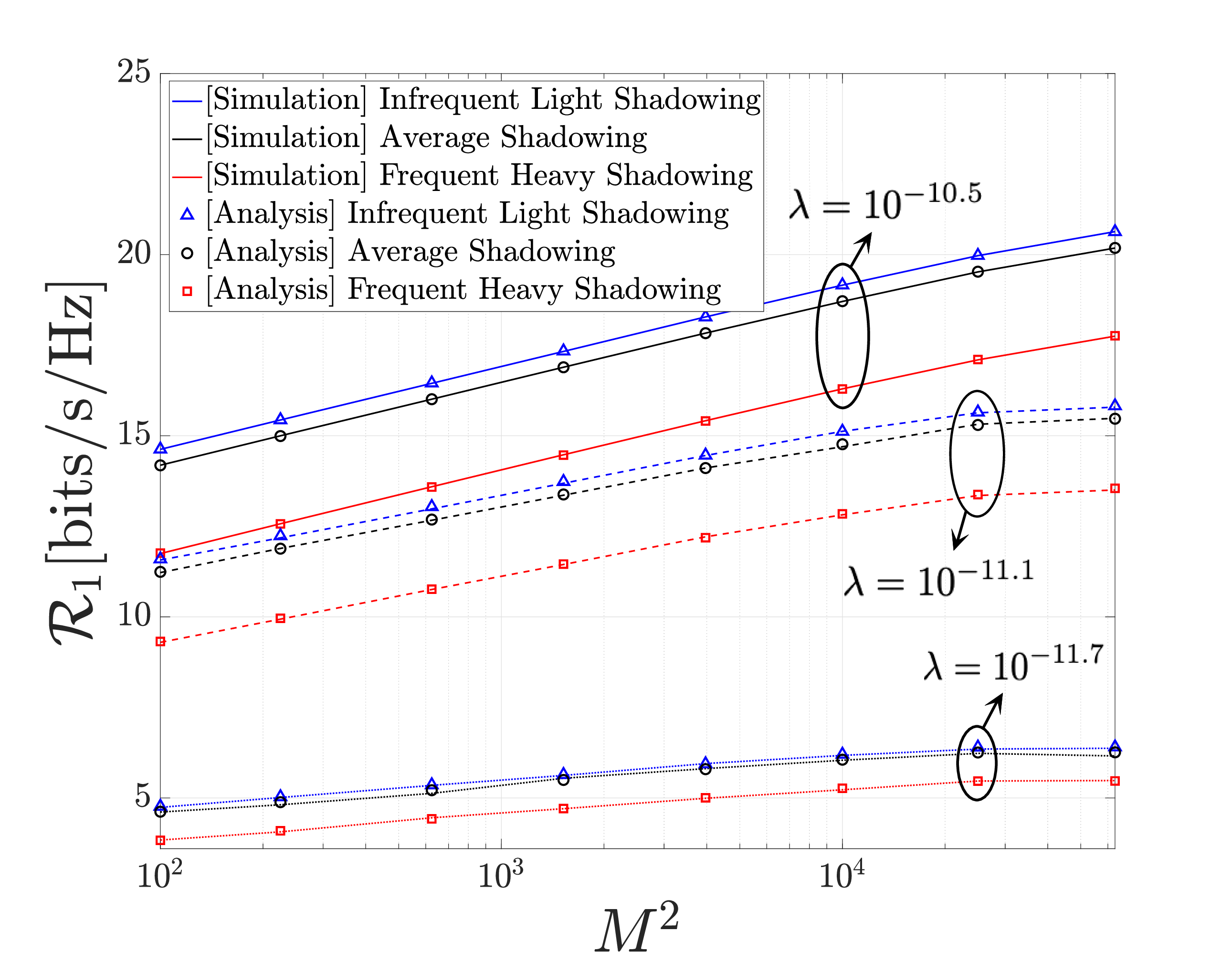}}} 
        \caption{In a single beam case, $\mathcal{R}_1$ versus $M^2$ with radius $R_1 = 250$km for different $\lambda$.}
        \label{fig:single beam}
\end{figure}
{\textcolor{black}{
We note that the ergodic rate has been widely adopted for evaluating throughput performance of mobile communications. Given that 3GPP non-terrestrial network (NTN) is considered to support the service of handheld terminals \cite{3gpp:NTN:2023}, it justifies our analysis on the ergodic rate in the considered GEO multibeam satellite system.}}
Now, we validate our analysis by comparing it with the simulation result. 
{\textcolor{black}{We note that the parameters used in the simulation are listed in Table \ref{table:SR} and Table \ref{table:parameter}. The simulation provides the distinct SR scenarios in Table \ref{table:SR} according to different $\lambda$.}} 
We also clarify that Table \ref{table:parameter} is referred to in \cite{zheng:twc:2012, koyoungcahi:twc:2022,wang:twc:19}. 
Fig.~\ref{fig:single beam} shows that $\mathcal{R}_1$ for different $\lambda$ as increasing $M^2$ with $R_1 = 250$km whose beam width is typically used in GEO satellite. The result indicates that the analytical results are well matched to the numerical simulations. 
An interesting observation of Fig.~\ref{fig:single beam} is that the scaling behavior of the ergodic rate $\mathcal{R}_1$ with $M^2$ is different depending on $\lambda$. That is to say, when $\lambda$ is sufficiently large, $\mathcal{R}_1$ increases with $M^2$, while $\lambda$ is relatively small, the growth of $\mathcal{R}_1$ rather slows down. 
In particular, when $\lambda = 10^{-11.7}$, increasing $M^2$ does not necessarily increase $\mathcal{R}_1$; but $\mathcal{R}_1$ rather decreases as $M^2$ increases.\footnote{
{\textcolor{black}{Unfortunately, when $M^2$ exceeds $10^5$, MATLAB encounters computational limitations that prevent the computation.}}}
The rationale behind this is as follows. 
Recall that we employ the fixed-beam precoding approach, in which the precoding vectors are not adjusted depending on CSI. 
As a result, it is possible that the selected user is not located at the exact beam center point, causing a beam mismatch. 
As illustrated in Fig.~\ref{fig:beam gain}, this beam mismatch results in a reduction of the beam gain. 
Now, let's assume that $M^2$ increases asymptotically. If the user is exactly at the beam center, the SNR also increases asymptotically due to the boosted array gain. 
On the contrary, if the beam mismatch occurs, increasing $M^2$ leads to narrower beam width; thereby the selected user tends to be located outside of the main beam width. 
For instance, $M^2 \rightarrow \infty$, the main-lobe beam width also goes to $0$ and this makes the corresponding beam gain $0$ for constant $\lambda$. 
To prevent this, $\lambda$ should scale up with $M$. 
{\color{black}{This observation is consistent with other studies that have considered fixed-beam precoding with user selection \cite{sharif:tit:05, gilwon:twc:2016}. This result highlights the importance of ensuring sufficient user density for the effective use of fixed-beams in practical scenarios.}}
In summary, to ensure a non-vanishing ergodic rate in multibeam satellite communication with massive MIMO where $M^2$ is very large, $\lambda$ should increase with $M$ at a certain scaling parameter, i.e., $\lambda \sim M^{q}$ (here, $x \sim f(M)$ implies $\lim_{M\rightarrow \infty}\frac{x}{f(M)}=1$). Identifying the scaling parameter $q$ is crucial in understanding and designing the considered satellite communication system. 


{\textcolor{black}{
Our scaling law analysis is also meaningful from an analytical perspective. That is to say, the ergodic rate obtained in \eqref{eq:raw R1} involves multiple integrals, making it challenging to evaluate the impact of each parameter on the ergodic rate. This complicated analytical expression is a common issue in the existing work that employed tools of stochastic geometry, such as \cite{koyoungcahi:twc:2022, park:twc:2023, kim:twc:24, jung:tcom:22, lim:twc:2024, talgat:taes:24}. 
To address this, the rate scaling law analysis provides a concise way to capture the interplay between key system parameters, such as user density and the number of antennas and beams. 
It is worth noting that, on the contrary to Corollary \ref{coro: single achievable rate}, which can be computed directly from \cite{hamdi:tcom:10}, the scaling law is obtained through our unique mathematical approach, representing one of this paper's key technical contributions. We detail this derivation in the next subsection.}}

\subsection{Asymptotical Scaling Law Analysis}
We now present the rate scaling law for the single beam case, which is one of the main results of paper.
{\textcolor{black}{\begin{theorem} \label{thm single r}
Let $\lambda \sim M^q$ for any $q \in (p+1+\epsilon/2, 2+\epsilon/2)$ and $p\in(0,1)$ with arbitrarily small $\epsilon > 0$.
Then, we have asymptotic upper and lower bounds of $\mathcal{R}_1$ as 
\begin{align}
    \log M^{2(q-1-\epsilon)} < \mathcal{ R}_1 < \log M^{2(q-1+\epsilon)}, \; \text{ for } M \rightarrow \infty. \label{eq:single rate}
\end{align}
\end{theorem}
\begin{proof}
Please see Appendix \ref{appendix:proof single r}.
\end{proof}}}
From the upper and lower bounds of Theorem \ref{thm single r}, we get $\mathcal{R}_1 \sim (q-1)\log M^2$. 
This implies that if $q > 1$, then the achievable ergodic rate $\mathcal{R}_1$ achieves a positive gain as $M \rightarrow \infty$. Otherwise $\mathcal{R}_1$ goes to $0$, i.e., it is infeasible to provide a stable ergodic rate in satellite communication with massive MIMO. 
Since the number of antennas on the UPA is $M^2$, $q=1$ corresponds to the square root of the number of antennas. This implies that the ground user density should scale with at least the square root of the number of UPA antennas. 
If $q = 2$, i.e., the user density scales at the same rate as the number of antennas, the achievable rate $\mathcal{R}_1$ scales with $\log M^2$, which indicates the ideal ergodic rate when perfect CSI is given to the satellite. 
In the following theorem, we further reveal this. 

{\textcolor{black}{
\begin{theorem}\label{thm:single fraction}
    For $\lambda \sim M^q$ with $q \in (p+1+\epsilon/2, 2+\epsilon/2)$ for $p\in(0,1)$ with arbitrarily small $\epsilon > 0$, we have 
   \begin{align} 
        \lim_{M \rightarrow \infty} \frac{\mathcal{R}_1}{ \mathbb{E} \left[\log\left(1 +  P G_{\text{Tx}} G_{\text{Rx}}  L_1 |g_1|^2 M^2 \right) \right] } = \begin{cases}
        q - 1, & \text{for } q \ge 1, \\
        0, & \text{for } q < 1.
        \end{cases}
        \nonumber
    \end{align}
\end{theorem}
\begin{proof}
    See Appendix \ref{appendix:proof coro single}.
\end{proof}}}



The denominator $\mathbb{E} \left[\log\left(1 +  P G_{\text{Tx}} G_{\text{Rx}} L_1 |g_1|^2M^2 \right) \right]$ in Theorem \ref{thm:single fraction} corresponds to the ideal ergodic rate by matching the beam center to the corresponding user's location, i.e., $f_{Z_1}(r_1,\phi_1) = 1$.
To this end, the selected user needs to send the CSI feedback to the satellite, and then the satellite aligns its precoding vector to the received CSI. 
Since the fixed-beam precoding approach does not adjust the precoding vector to the ground user, 
the ideal ergodic rate is consistently larger than the achievable ergodic rate $\mathcal{R}_1$. 
For this reason, the ratio in Theorem \ref{thm:single fraction}
is interpreted as the extent of performance degradation caused by not sending the CSI feedback. 
The fixed-beam precoding approach achieves the fraction of $q-1$ of the ideal rate when $\lambda \sim M^q$. 
From this, we find that $q>1$ is necessary to achieve a nonvanishing ergodic rate, as observed in Theorem \ref{thm single r}. If $q = 2$, i.e., $\lambda \sim M^2$, then the fixed-beam precoding approach asymptotically achieves the ideal ergodic rate, implying that no CSI is needed to achieve the optimal rate. 



\section{A Multiple Beam Case}\label{sec:multibeam} 
In this section, we extend our analysis by incorporating a multiple beam case. 
We consider that the satellite forms $K$ number of beams to serve $K$ spot regions. 
For the $k$-th spot region $\CMcal{A}_k$, we select a user according to \eqref{eq:user_sel} and use the precoder as described in \eqref{eq:precoding vector}. 
Without loss of generality, we denote the user index selected for beam $k$ as $k$
and beam $1$ is the nadir-pointing located at the center of the coverage region. 
Similar to the single beam case, we first characterize the achievable ergodic rate as a function of the system parameters and study the scaling laws. 


\subsection{Achievable Rate Analysis}
In the multibeam scenario, it is of importance to properly account for the amount of inter-beam interference. 
To this end, we denote \textcolor{black}{$f_{Z_{i}}(r_k,\phi_k) =  |\mathbf{h}_k^{\sf H} \mathbf{f}_{i}|^2$} as the beam gain that user $k$ receives from the $i$-th fixed-beam. Accordingly, $f_{Z_i}(r_k,\phi_k)$ indicates the amount of interfering beam gain from the $i$-th beam for $i \neq k$. With this, the signal-to-interference-plus-noise-ratio (SINR) of user $k$ is given by
\begin{align}\label{eq:SINR}
    \text{SINR}_k = \frac{ \bar{P} G_{\text{Tx}} G_{\text{Rx}} L_k |g_k|^2 M^2 f_{Z_{k}}(r_k,\phi_k)  }{ \bar{P} G_{\text{Tx}} G_{\text{Rx}} L_k |g_k|^2 \sum_{i\neq k} M^2 f_{Z_{i}}(r_k,\phi_k)  + 1},
\end{align}
where the allocated transmit power $P_0$ is divided by $K$ as $\bar{P} = P/K$, $L_k$ is the large-scale fading of user $k$ as in \eqref{eq:LSF} and $|g_k|^2$ is the fading power drawn from the PDF \eqref{eq:SSF}.
The sum ergodic rate of multiple beam case is defined by
\begin{align}\label{eq:sum rate}
    \mathcal{R}_{\Sigma} = \sum_{k=1}^K \mathcal{R}^M_k =   \sum_{k=1}^K \mathbb{E}\left[ \log\left(1+\text{SINR}_k \right) \right].
\end{align}
Now we analyze the achievable ergodic rate in the multibeam case. 
In this analysis, we focus on user $1$'s ergodic rate as a representative case. 
{\color{black}{The rationale for choosing the nadir-pointing beam as a representative case is as follows:  
First, since beam $1$, which is a nadir-pointing beam as shown in Fig.~\ref{fig:system model}, is surrounded by other beams, user $1$ is most susceptible to inter-beam interference. This represents the worst-case scenario and serves as the lower bound for the network performance. 
Second, if extending to large networks with multiple satellites that each cover their own area, users at the edge of each coverage may experience inter-satellite interference. Although we analyze a single satellite here, it is reasonable to investigate the performance of the nadir-pointing beam using the wrap-around technique to understand the overall system performance. 
We derive the ergodic rate in the multibeam case as follows. }}
    \begin{figure*}
    \begin{align}
        \mathcal{R}^M_1
        &\mathop{=} \int_0^R \int_0^{2\pi} \int_0^\infty \frac{1}{\tau} \left( \frac{(2b_0m)^m \left(1+2 \tau b_0 \bar{P} G_{\text{Tx}} G_{\text{Rx}} L_1 
        \sum_{i \neq 1} M^2 f_{Z_{i}}(r,\phi) \right)^{m-1}}{\left[(2b_0m + \Omega)\left(1+2 \tau b_0  \bar{P} G_{\text{Tx}} G_{\text{Rx}}L_1 
        \sum_{i \neq 1} M^2 f_{Z_{i}}(r,\phi) \right)-\Omega \right]^m }  \right. \nonumber  \\ 
        & \left. - \frac{(2b_0m)^m \left(1+2 \tau b_0 \bar{P} G_{\text{Tx}} G_{\text{Rx}} L_1 
        \sum_{i =1 } M^2 f_{Z_{i}}(r,\phi) \right)^{m-1}}{\left[(2b_0m + \Omega)\left(1+2 \tau b_0 \bar{P} G_{\text{Tx}} G_{\text{Rx}} L_1 
        \sum_{i =1 } M^2 f_{Z_{i}}(r,\phi) \right)-\Omega \right]^m }  \right)e^{-\tau} \lambda r e^{-\pi \lambda r^2}  d\tau d\phi dr  \label{eq:rate k in MU}
    \end{align}
\noindent\hrulefill 
\end{figure*}
{\textcolor{black}{
\begin{corollary}\label{coro:MU single beam}
    In the multibeam case, we define the ergodic rate of user $1$ as
    \begin{align}
        \mathcal{R}^M_1 = \mathbb{E}\left[ \log \left( 1 + \frac{ \bar{P} G_{\text{Tx}} G_{\text{Rx}} L_1 |g_1|^2 
        M^2 f_{Z_1}(r_1,\phi_1)  }{ \bar{P} G_{\text{Tx}} G_{\text{Rx}} L_1 |g_1|^2 \sum_{i \neq 1} M^2 f_{Z_{i}}(r_1,\phi_1)  + 1} \right) \right], \nonumber
    \end{align}
    where the expectation is about the randomness associated with the fading power, spatial locations of ground users. Then, the $\mathcal{R}^M_1$ is obtained in \eqref{eq:rate k in MU}.
\end{corollary}
\begin{proof}
    See Appendix \ref{appendix:MU single beam}.
\end{proof}}}
\begin{figure}[t]
    \centerline{\resizebox{1\columnwidth}{!}{\includegraphics{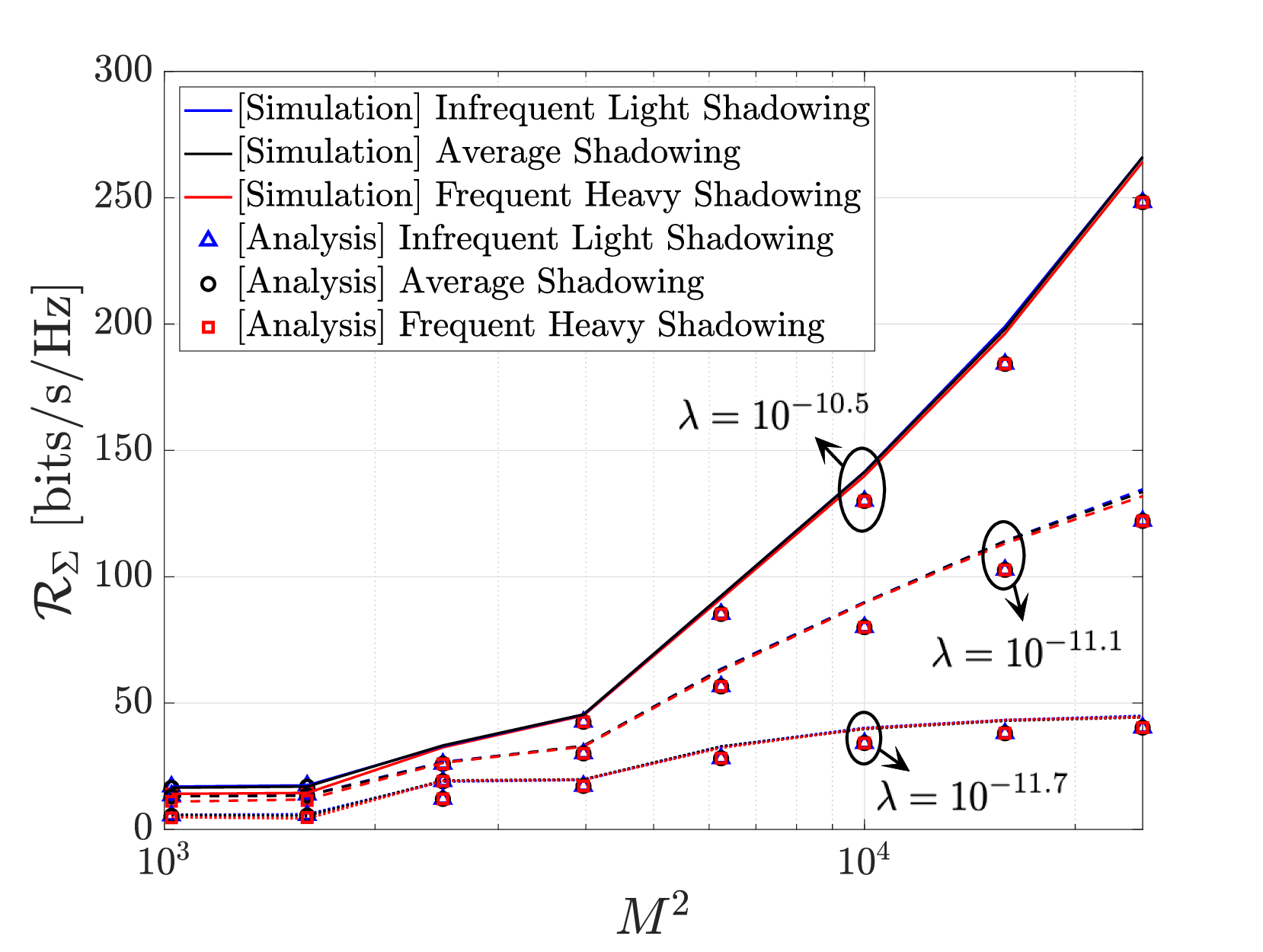}}} 
        \caption{In a multiple beam case, \textcolor{black}{$\mathcal{R}_\Sigma$} versus $M^2$ for different $\lambda$. $K$ is determined as the number of beams that completely fill the whole coverage area with the beam configuration in \eqref{eq:beam position} with $\ell=1$.}
        \label{fig:multibeam sumrate}
\end{figure}
Fig.~\ref{fig:multibeam sumrate} shows the $\mathcal{R}_\Sigma$ versus $M^2$ for different parameters $\lambda$ and SR. 
In the analysis, all beams are assumed to be nadir-pointing and surrounded by interfering beams. Accordingly, the analysis results are obtained by multiplying $\mathcal{R}^M_1$ in \eqref{eq:rate k in MU} by the total number of beams $K$. 
In contrast, in the simulation, some beams are located at the coverage edge and thus experience less interference. 
This leads to discrepancies between the analysis and simulation, with the analysis serving as the lower bound, as observed in Fig.~\ref{fig:multibeam sumrate}.
The parameters used in Fig.~\ref{fig:multibeam sumrate} are listed in Tables \ref{table:SR} and \ref{table:parameter}.
{\textcolor{black}{The curve in Fig.~\ref{fig:multibeam sumrate} is not smooth because the number of beams is determined by $M^2$ based on the multibeam configuration in \eqref{eq:beam position}.}}


Understanding the asymptotic behavior between the user density $\lambda$, the number of beams $K$, and the number of antennas $M^2$ is crucial for gaining design insight into the multibeam satellite communication system. 
However, analyzing the multiple beam case is more complicated compared to the single beam case 
because of the challenge of capturing inter-beam interference. 
In the next subsection, we clarify the difficulty and put forth our idea to resolve this. 


\subsection{Asymptotical Scaling Law Analysis}

In this subsection, we study the scaling laws between $\lambda$, $K$, and $M^2$. 
A key hindrance of the analysis is characterizing the amount of inter-beam interference in a tractable manner. 
The amount of inter-beam interference is mainly determined by the inter-beam spacing and the beam width. 
To capture this, we recall that the inter-beam spacing is controlled by the parameter $\ell$ as outlined in the beam configuration \eqref{eq:beam position}, wherein we examine within the range $\ell \in (0,1)$ in the analysis.
It is clear that increasing $\ell$ narrows the inter-beam spacing, leading to higher inter-beam interference. However, this allows for more beam multiplexing gains to be attained by using more beams. 
Conversely, decreasing $\ell$ alleviates the inter-beam interference, while limiting the beam multiplexing gains. 


\begin{figure}[t]
    \centerline{\resizebox{1\columnwidth}{!}{\includegraphics{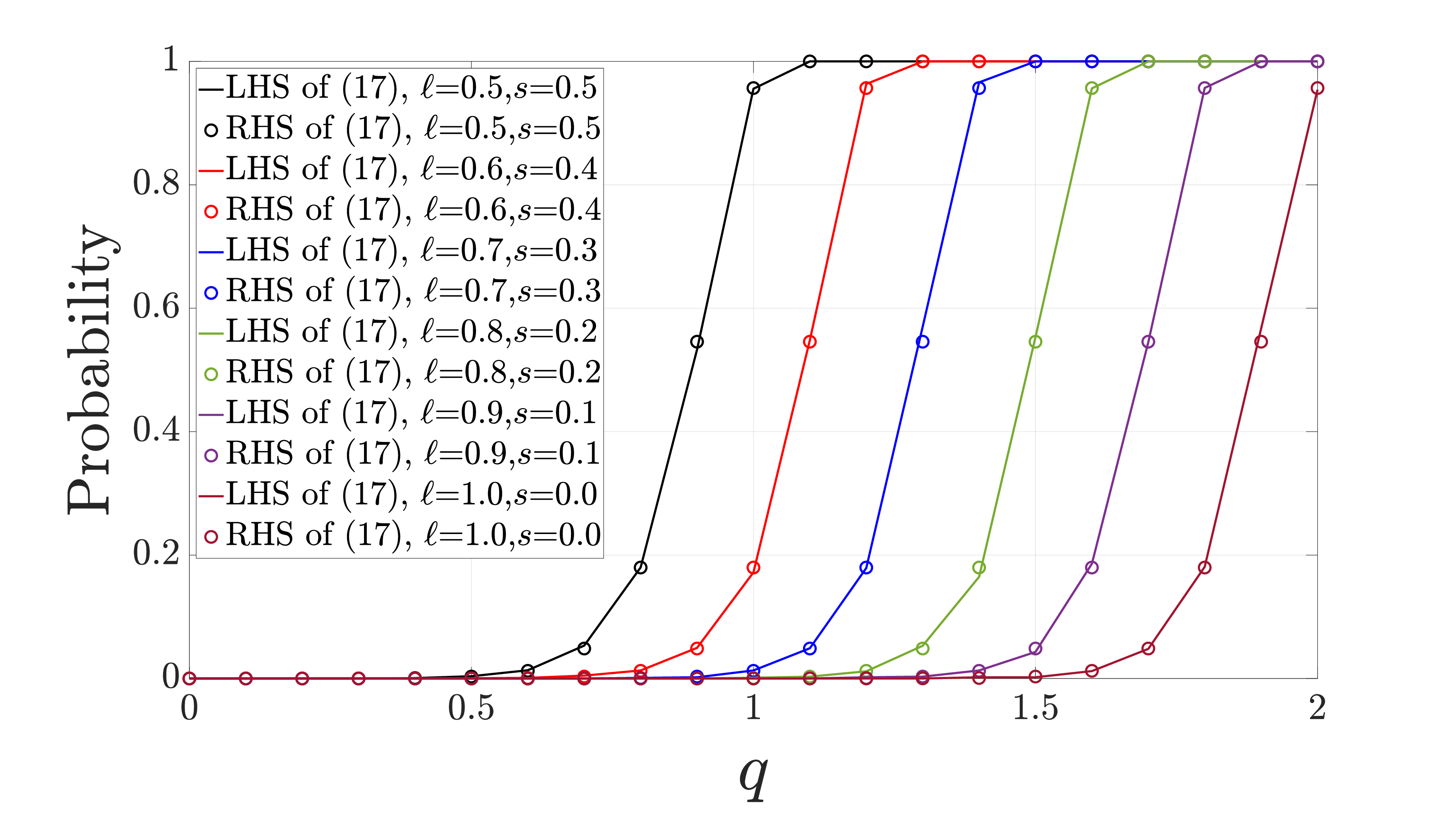}}} 
        \caption{In a multiple-beam case, the simulation results of \eqref{eq:result interf} as increasing $q$ under different sets of $\ell$ and $s$.}
        \label{fig:interference}
\end{figure}
As a key ingredient of the scaling law analysis in the multiple beam case, 
we comprehend the inter-beam interference experienced by the user $1$ from another fixed-beam in relation to $\ell$. 
Since the number of spot beams that can be packed in $\CMcal{A}$ is scaled with $M^{2\ell}$, 
we assume $K = K_0 M^{2\ell}$ where $K_0$ is obtained by solving a circle packing problem for given inter-beam spacing. 
Obtaining $K_0$ for specific $\ell$ and $\CMcal{A}$ is interesting yet beyond the scope of our paper. 
{\color{black}{
In the asymptotic regime of our interest, it is possible to choose $K_0=1$ because the inter-beam spacing becomes sufficiently small as $M$ increases, so that $M^{2\ell}$ number of beams can be packed within $\CMcal{A}$.}}
We characterize the inter-beam interference in the following lemma. 
\begin{lemma}\label{lem:single interf}
    For $s,\ell \in (0,1)$ such that $\ell + s <1$, we have 
    \begin{align}
    \mathbb{P}\left[ \frac{M^2}{M^{2\ell}} f_{Z_i}(r_1,\phi_1) < \frac{1}{M^{2s}}\right] > 1 - \exp\left[ - \frac{\lambda \pi H^2}{M^{2\ell}-1 }  \right]  \label{eq:result interf}
    \end{align}
    for $M \rightarrow \infty$.
\end{lemma}
\begin{proof}
Please see Appendix \ref{proof of lem:single interf}.
\end{proof}
Fig.~\ref{fig:interference} shows the probability that the interference from an adjacent beam (where $n=1$ and $m=1$ in \eqref{eq:beam position}) is below a certain level, i.e., $\frac{1}{M^{2s}}$. 
As shown in Fig.~\ref{fig:interference}, the left-hand-side (LHS) of Lemma \ref{lem:single interf} is followed by the right-hand-side (RHS) of Lemma \ref{lem:single interf}. The results validate the result of Lemma \ref{lem:single interf}.

To explore Lemma \ref{lem:single interf} deeply, we introduce an auxiliary variable $\delta$ such that $\ell + s + \delta = 1$. 
If $\ell$ increases, then $s$ decreases. In this case, the beam spacing becomes narrow, which leads to an increase in interference level. 
Nevertheless, to ensure the interference remains to be below certain level, more user density is required to compensate for the reduced beam spacing according to $\ell$, which is denoted as $q>2\ell$.
Conversely, if $\ell$ decreases, then $s$ increases, which means the beam spacing widens, allowing for less interference. 

\textcolor{black}{Lemma \ref{lem:single interf} provides insight into why selecting the user closest to each beam is effective in a multibeam environment. Specifically, the array gain in the numerator of $\text{SINR}_1$ remains identical to that of the single beam case (i.e., $\text{SNR}_1$), which makes location-based user selection an effective means of increasing the desired signal power. Moreover, Lemma 1 shows that the channel of the user closest to the beam center exhibits asymptotically favorable propagation.
It is worth noting that our approach may not be globally optimal in general. Nonetheless, it is effective in multibeam satellite communications, particularly considering the inherent constraints of such systems. Furthermore, Theorems \ref{thm:single interf rate} and \ref{thm:sum rate} demonstrate that this approach can achieve optimal performance in a specific operational region.}
{\textcolor{black}{
\begin{theorem}\label{thm:single interf rate}
    Let $\lambda \sim M^q$ with $p, \ell \in (0,1)$ and $\epsilon>0$ such that $q \in (p+1+\epsilon/2, 2+\epsilon/2)$, $\ell + s < 1$ and $q > 2\ell$. For $M\rightarrow \infty$, the expected rate of user $1$ with multiple beam is given by
    \begin{align}
    \log M^{2(q-\ell - 1-\epsilon)} < \mathcal{R}_1^M < \log M^{2(q-\ell - 1 + \epsilon)}.
    \end{align}
\end{theorem}
\begin{proof}
    Please see Appendix \ref{appendix:single interf rate}.
\end{proof}}}
Now we elucidate Theorem \ref{thm:single interf rate}. To achieve non-vanishing performance in multiple fixed-beam, a larger density $\lambda$ is required such that $q>\ell + 1$, which contrasts to the single beam case which requires $q>1$.
In other words, in multibeam scenarios, it is necessary to increase the user density by $\ell$ to compensate for the impact of interference. 
We extend Theorem \ref{thm:single interf rate} to the sum rate in the following theorem. 
{\textcolor{black}{\begin{theorem}\label{thm:sum rate}
    For $\lambda \sim M^q$ with $p,\ell\in (0,1)$ such that $q \in (p+1+\epsilon/2, 2+\epsilon/2)$, $\ell+s<1$ and $q > 2\ell$, we have
    \begin{align}
        \lim_{M\rightarrow \infty} \frac{\mathcal{R}_\Sigma}{ K\cdot \mathbb{E}\left[ \log \left(1 + \bar{P} G_{\text{Tx}} G_{\text{Rx}} L_1 |g_1|^2 M^2  \right) \right] } \nonumber \\=
        \begin{cases}
         \frac{q-\ell-1}{1-\ell}, & \text{for } q \ge \ell + 1, \\
         0, & \text{for } q < \ell + 1.
        \end{cases}
    \end{align}
\end{theorem}
\begin{proof}
    Please refer to Appendix \ref{appendix:coro sum rate}.
\end{proof}}}
The denominator $\mathbb{E}\left[ \log \left(1 + \bar{P} G_{\text{Tx}} G_{\text{Rx}} L_1 |g_1|^2 M^2  \right) \right]$ denotes the ideal ergodic rate for user $1$ by perfectly eliminating interference, provided that precoding is used with perfect CSI. 
To be specific, ${q-\ell-1}>0$ implies that the considered fixed-beam precoding method achieves such a fraction of the optimal performance, while $\frac{1}{1-\ell}>1$ for $\ell \in (0,1)$ implies the multiplexing gain. 
From this perspective, when focusing on a single beam among multiple beams, an additional user density of $\ell$ is required compared to a single beam scenario without interference. However, with the additional required $\ell$, a multiplexing gain of $\frac{1}{1-\ell}$ can be achieved. \textcolor{black}{Notably, for the user density scaling with the number of antennas, i.e., $q=2$, the proposed user selection strategy achieves the asymptotically optimal sum rate for multibeam satellite communications.} On the other hand, when the number of beam $K=1$, i.e., $\ell=0$, the result of Theorem \ref{thm:single fraction} is reduced to $q-1$, which matches well with the result of Theorem \ref{thm:single fraction} for a single beam case.
Furthermore, the sum rate exhibits different slopes with respect to $\lambda$ as $M^2$ increases. This aligns with the results with Fig.~\ref{fig:multibeam sumrate}.

{\textcolor{black}{
\begin{remark} [Guidelines for practical satellite network design] \normalfont
    In practical multibeam satellite communication systems, increasing user density can be challenging. Nevertheless, our analysis still offers valuable guidance for real-world satellite network deployments. Specifically, we derive the rate scaling law and demonstrate that the interactions between the number of antennas and user density can offset each other. This provides insights into how user density impacts the achievable rate and how it can be compensated by adjusting other system parameters. By leveraging these relationships, satellite operators can optimize key resources, such as the number of antennas, beams, and bandwidth. For instance, given a specific user density, the satellite network can predict the achievable rate and determine the required bandwidth to meet quality-of-service (QoS) requirements. Additionally, our analysis helps in adjusting the number of beams or beam spacing to balance multiplexing gains with inter-beam interference. This allows for efficient network design, ensuring performance enhancement for varying operational scenarios.
\end{remark}
}}

\begin{figure}[t]
    \centerline{\resizebox{1\columnwidth}{!}{\includegraphics{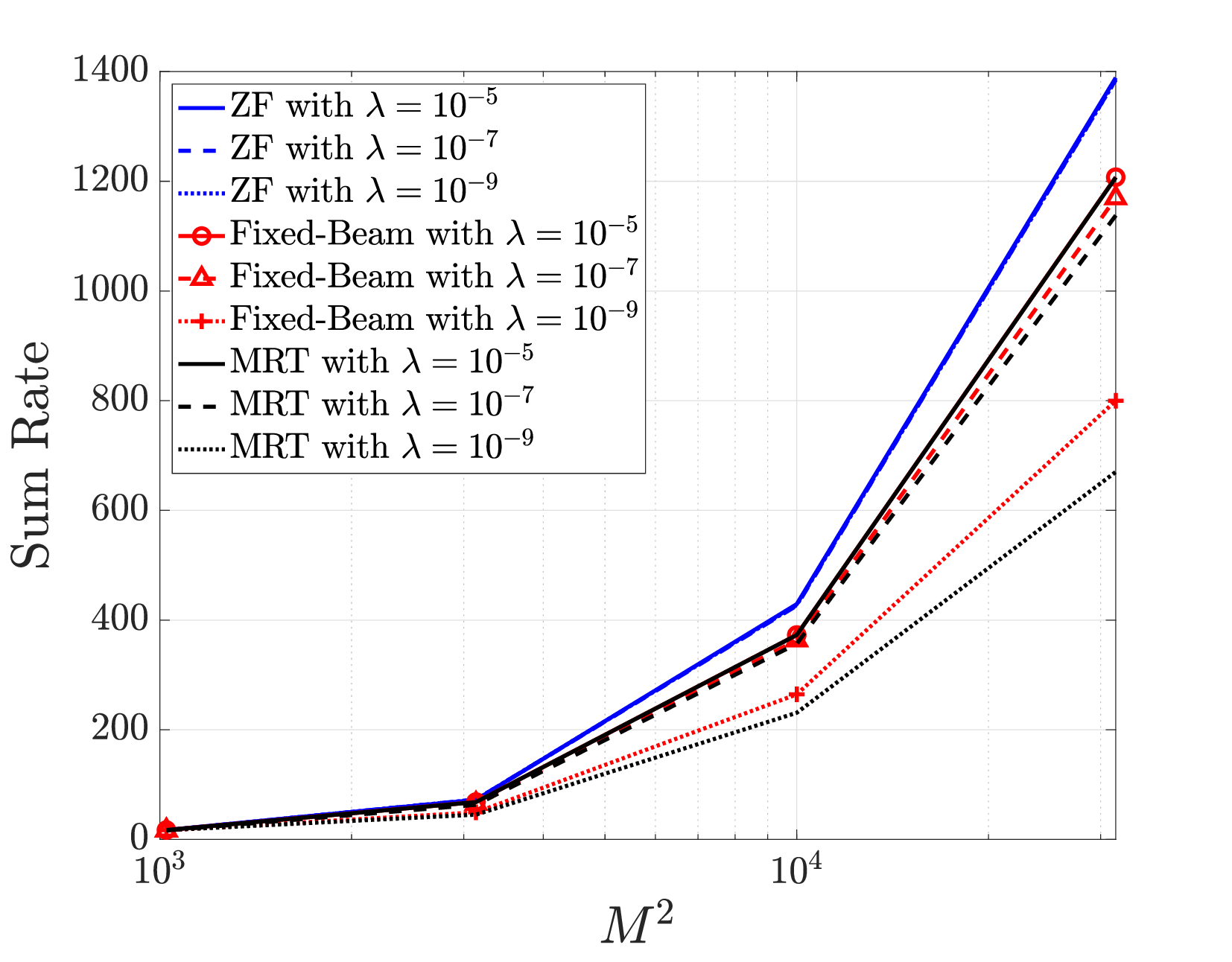}}} 
        \caption{In multiple beam case, the sum rate versus $M^2$ for different user densities $\lambda$. $K$ is determined as the number of beams that completely fill the whole coverage area with the beam configuration in \eqref{eq:beam position} with $\ell=1$. In the simulation, MRT and ZF precoders are computed for the same set of users selected by our fixed-beam precoding method via \eqref{eq:user_sel}, under the assumption of perfect CSI.}
        \label{fig:ZFMRT}
\end{figure}

{\color{black}
{\begin{remark}[Comparison with MRT and ZF] \normalfont 
    We compare the proposed fixed-beam precoding approach with two widely studied MIMO precoding techniques: maximum ratio transmission (MRT) and zero-forcing (ZF).
    To do this, we assume a transmitter equipped with $M$ antennas serves $K$ single-antenna users. MRT is the simplest linear precoding technique, designed to maximize the received SNR by aligning the precoding vector with the user's channel. It directly employs the conjugate transpose of the channel vector as the precoder. Applying MRT precoder to the message vector requires $M$ multiplications and additions. Thus, its computational complexity is $\mathcal{O}(KM)$. MRT demands updating the precoder whenever the instantaneous CSI changes. In addition, MRT does not account for inter-user interference and thus cannot support spatial multiplexing, which limits its effectiveness in multiuser scenarios. In contrast, ZF is a linear precoding method that provides spatial multiplexing by actively eliminating inter-user interference. This is achieved by projecting each user's signal onto the null space of the interference, typically through matrix inversion. The computational complexity of ZF is $\mathcal{O}(K^3 +K^2M + KM)$.
    
    Despite these existing precoding methods, they are generally unsuitable for practical multibeam satellite communications. 
    The rationale behind this is as follows: i) It is challenging to obtain accurate CSI in satellite systems; ii) The hardware constraints of satellites, including limited onboard computational power, size, weight, and energy, prohibit complex precoding computations such as matrix inversion; iii) Joint gateway processing is typically required for centralized processing, which introduces additional delays and demands considerable bandwidth for CSI exchange between gateways. For the above reasons, 3GPP NTN \cite{3gpp:NTN:2023} has adopted fixed-beam precoding in GEO satellite communications, where the precoder is fixed toward a specific point and does not adapt to instantaneous CSI changes. As a result, the fixed-beam precoder does not require accurate CSI. Instead, it only requires applying precoders to the message vectors, resulting in a computational complexity of $\mathcal{O}(KM)$. This approach avoids matrix inversion and eliminates the need for joint gateway processing. Additionally, configuring each beam toward a null point effectively mitigates interference.
    
    Fig.~\ref{fig:ZFMRT} compares the sum rate performance of proposed fixed-beam precoding with those of MRT and ZF for different user densities $\lambda$. In the simulation, ZF and MRT precoders are also computed for the same set of users selected by the fixed-beam precoding method according to \eqref{eq:user_sel}, under the assumption of perfect CSI. The result shows that as user density $\lambda$ increases, the performance of fixed-beam precoding approaches that of ZF. This demonstrates that our user selection effectively mitigates interference in large $\lambda$ regime. Furthermore, the performance of MRT and the proposed fixed-beam approach becomes similar as $\lambda$ grows. Although MRT relies on full CSI, the proposed fixed-beam precoding achieves comparable performance without requiring instantaneous CSI from ground users. MRT typically fails to achieve spatial multiplexing; however, the semi-orthogonal user channels obtained by our user selection \eqref{eq:user_sel} enables MRT to achieve it, as shown in Fig.~\ref{fig:ZFMRT}.

    Although ZF achieves higher rate performance, fixed-beam precoding clearly stands out by offering negligible computational complexity and requiring no instantaneous CSI. Instead, only the location information of each user is required. MRT also has low computational complexity; however, fixed-beam precoding is more suitable for multibeam satellite communications due to its effective interference mitigation and independence from instantaneous CSI.

    In summary, while fixed-beam precoding may not be globally optimal in general, it effectively mitigates interference with low computational complexity by designing each beam to reflect the inherent characteristics of satellite communication systems. We further provide analytical results in Theorems \ref{thm:single fraction} and \ref{thm:sum rate}, demonstrating that fixed-beam precoding achieves optimal performance in a specific operational regime. This constitutes our main theoretical contribution.
\end{remark}}}

\section{Conclusion}
In this paper, we have considered a fixed-beam precoding approach for massive MIMO multibeam satellite communication systems combined with a location-based user selection strategy. 
Upon this, we have provided a performance analysis that sheds light on the asymptotical interplay between the density of ground users, the number of beams, and the number of antennas. 
Our major findings are that when the user density is scaling at the identical rate as the number of antennas, then the fixed-beam precoding is able to provide enough beam gain even without CSI while the beam mismatch becomes negligible. 
In the multiple beam case, we have found that the scale of the interference is adjusted by the beam spacing, while providing the probability of the interference scale as a function of user density. Moreover, the fixed-beam precoding, when user density scales with the number of antennas, achieves the asymptotic optimal sum rate regardless of beam spacing.
The current analysis is based on a single user per beam approach, and extending this analysis to a multicast scenario remains future work.


\appendices
\section{Proof of Corollary \ref{coro: single achievable rate}} \label{appendix: single achievable rate}
Assuming a random variable $X=|g_1|^2$ where $g \sim \CMcal{SR}(\Omega, b_0, m)$, we have 
    \begin{align}
        &\mathbb{E}[\log(1+Xh(r_1,\phi_1))] \notag \\
        & \mathop{=} \mathbb{E}[\mathbb{E}[\log(1+Xh(r_1,\phi)) |r_1,\phi_1]] \notag \\
        &\mathop{=}^{(\text{a})} \mathbb{E}\left[ \int_0^\infty \frac{1}{\tau} \left(1-\mathbb{E}\left[e^{-\tau  h(r_1,\phi_1) X } \right] \right) e^{-\tau} d\tau \middle| r_1,\phi_1 \right] \notag \\
        &\mathop{=}^{(\text{b})} \int_0^{2\pi} \int_0^R \int_0^\infty \frac{1}{\tau} \left( 1 - \frac{(2b_0m)^m (1+2 \tau b_0 h(r,\phi) )^{m-1}}{[(2b_0m + \Omega)(1+2 \tau b_0 h(r,\phi))-\Omega]^m } \right)  \notag \\
        & \phantom{----------} \phantom{------------} e^{-\tau} \lambda r e^{-\pi \lambda r^2} d\tau d\phi dr   \notag 
    \end{align}
    where (a) follows \cite{hamdi:tcom:10}
    \begin{align}
        \log(1+x)= \int_0^\infty \frac{1}{\tau}(1-e^{-\tau x})e^{-\tau}d\tau
    \end{align}
     and (b) follows the moment generating function (MGF) of $X$ given by \cite{abdi:twc:2003}
    \begin{align}
        \mathbb{E}\left[e^{-s X} \right] = \frac{(2b_0m)^m (1+2b_0 s)^{m-1}}{[(2b_0m + \Omega)(1+2b_0s)-\Omega]^m } \label{eq:SR mgf}.
    \end{align}
    Now we obtain the PDF of $r_1$ and $\phi_1$. 
    Recalling that $\Phi_1 = \{{\bf{d}}_i \in \CMcal{A}_1\}$, we get the conditional PDF of $r_1$ as 
    \begin{align} 
        f_{r_1|\Phi_1>0}(r_1) =
        \begin{cases}
        \frac{2\lambda \pi r e^{-\lambda \pi r^2}}{1-e^{-\lambda \pi R_1^2}}, &\text{$ 0 \le r_1 \le R_1$} \label{eq:nearest dist},  \\
        0 & \text{otherwise}.
        \end{cases} 
    \end{align}
    The proof of \eqref{eq:nearest dist} is straightforward in proof of Lemma \ref{lem:prob distance}, especially \eqref{eq:appendix1 eq 1}. 
    Here, $r_1$ is independent to $\phi_1$ which is calculated in isolation by $\phi \sim \text{Unif}[0,2\pi]$. Then, by putting $h(r,\phi) = P G_{\text{Tx}} G_{\text{Rx}} L_1 M^2 f_{Z_1}(r, \phi)$, this completes the proof.

\section{Proof of Theorem \ref{thm single r}}\label{appendix:proof single r}
Before proving Theorem \ref{thm single r}, we introduce the useful lemma to use the subsequent proofs. 
\begin{lemma} \label{lem:prob distance}
We denote the homogeneous PPP $\Phi_i = \{\mathbf{d}_1,\cdots,\mathbf{d}_{N_i} \}$ where $N_i$ follows the PPP with average number $\lambda \pi R_i^2$. Then, the probability that the distance of the nearest user from the nadir-pointing beam $r$ is in range between $R_a$ and $R_b$ is given by
    \begin{align}
        \mathbb{P}\left[R_a < r < R_b \right] =\exp\left[-\lambda \pi R_a^2\right] - \exp\left[-\lambda \pi R_b^2  \right] \label{appendix:lemma2:prob r}
    \end{align}
    where $0\leq R_a \leq R_b$.
\end{lemma}
\begin{proof}
The probability of $r$ which is the distance from the beam to nearest user within the range between $R_a$ and $R_b$ where $0\leq R_a \leq R_b$ is given by
\begin{align}
    &\mathbb{P}[R_a < r < R_b] \nonumber \\
    &=\mathbb{P}\left[R_a < r < R_b \middle| \Phi_b > 0  \right]\mathbb{P}[\Phi_b>0] \notag.
\end{align}
For the case $\Phi_b=0$, the probability about $r$ is equal to 0. Then, we have
\begin{align}
    \mathbb{P}&\left[R_a < r < R_b  \middle| \Phi_b > 0  \right] \nonumber \\
    &= \mathbb{P} \left[\Phi_a = 0 | \Phi_b > 0 \right] 
    = \frac{\mathbb{P}[\Phi_a=0] (1- \mathbb{P}[ \Phi_b/\Phi_a  =0 ])}{\mathbb{P}[\Phi_b>0]} \notag \\
    &= \frac{\exp[-\lambda \pi R_a^2](1-\exp[-\lambda \pi (R_b^2 - R_a^2) ])}{\mathbb{P}[\Phi_b >0]}\notag \\
    &= \frac{\exp[-\lambda \pi R_a^2]-\exp[-\lambda \pi R_b^2 ]}{\mathbb{P}[\Phi_b>0]} \label{eq:appendix1 eq 1}
\end{align}
where $\Phi_b/\Phi_a$ is the independent PPP for $\CMcal{A}_b$ excluding the region of $\CMcal{A}_a$.
This is the end of the proof.
\end{proof}
To prove Theorem \ref{thm single r}, we investigate $\mathcal{\bar{R}}_{r}$ defined as
\begin{align} \label{eq:R r}
    \mathcal{\bar R}_{r} = \mathbb{E}_{r_1|\phi_1}\left[ \log\left(1 + \frac{1}{r_1^2 + H^2} Z \right)  \middle| \phi_1 \right],
\end{align} 
where $Z = M^2f_{Z_1}(r_1,\phi_1)$.
To do this, we analyze the event $\{Z > M^{2p} | \phi_1 \}$ conditioned on two case for $p \in (0,1)$: when $\phi_1 \neq 0$ and when $\phi_1=0$.

{
\textcolor{black}{
{$\bullet $ Case 1 ($\phi_1 \neq 0$)}: The event $\{ Z > M^{2p} \}$ is given by
\begin{align}
    \frac{1}{M^2} \left| \frac{ \sin\left(\frac{\pi M}{2} \sin\theta_1 \cos\phi_1 \right) \sin\left(\frac{\pi M}{2} \sin\theta_1 \sin\phi_1 \right)}
    { \sin\left(\frac{\pi }{2} \sin\theta_1 \cos\phi_1 \right) \sin\left(\frac{\pi }{2} \sin\theta_1 \sin\phi_1 \right)} \right|^2 > M^{2p} \notag
\end{align}
which is equal to
\begin{align}
     \left| \frac{ \sin\left(\frac{\pi M}{2} \sin\theta_1 \cos\phi_1 \right) \sin\left(\frac{\pi M}{2} \sin\theta_1  \sin\phi_1 \right)}
    { \sin\left(\frac{\pi }{2} \sin\theta_1 \cos\phi_1 \right) \sin\left(\frac{\pi }{2} \sin\theta_1 \sin\phi_1 \right)} \right| > M^{p+1}. \notag
\end{align}
The sufficient conditions for $\{ Z > M^{2p} | \phi_1 \}$ for small $\epsilon > 0$ are given by
\begin{align} \label{eq:suff1}
    \left|  \sin\left(\frac{\pi M}{2} \sin\theta_1 \cos\phi_1 \right) \sin\left(\frac{\pi M}{2} \sin\theta_1 \sin\phi_1 \right) \right| > \frac{1}{M^{\epsilon/2}}
\end{align}
and
\begin{align}\label{eq:suff2}
    \left| \sin\left(\frac{\pi }{2} \sin\theta_1 \cos\phi_1 \right) \sin\left(\frac{\pi }{2} \sin\theta_1 \sin\phi_1 \right) \right| < \frac{1}{M^{(p+1)+\epsilon/2}}.
\end{align}
Using the fact that $|\sin x| > \frac{|x|}{2}$ for small $x$, the sufficient conditions for $\eqref{eq:suff1}$ are
\begin{align}
    \frac{1}{2} \left| \frac{\pi M}{2} \sin\theta_1 \cos\phi_1 \right| \cdot \frac{1}{2}\left| \frac{\pi M}{2} \sin\theta_1 \sin\phi_1 \right| > \frac{1}{M^{\epsilon/2}} \notag
\end{align}
which can be reformulated as
\begin{align}
    | \sin \theta_1 | > \frac{1}{ \frac{\pi}{4} \sqrt{|\sin\phi_1 \cos\phi_1|} M^{1+\epsilon/4}}. \notag
\end{align}
Also, by using the fact $\sin x < x$ for $x \in (0,\pi/2)$, the sufficient condition for \eqref{eq:suff2} is denoted as
\begin{align}
    \left|\frac{\pi}{2} \sin\theta_1 \cos\phi_1 \right| \cdot \left| \frac{\pi}{2} \sin\theta_1 \sin\phi_1 \right| < \frac{1}{M^{(p+1)+\epsilon/2}} \notag
\end{align}
which is reformulated as
\begin{align}
    |\sin \theta_1| < \frac{1}{\frac{\pi}{2} \sqrt{|\sin\phi_1 \cos\phi_1|} M^{(p+1)/2 + \epsilon/4}}. \notag
\end{align}
Thus, the sufficient condition for the event $\{ Z > M^{2p} |\phi_1 \}$ is given by
\begin{align} \label{eq:range sin theta}
    \frac{1}{ \alpha M^{1+\epsilon/4}} < |\sin\theta_1| < \frac{1}{2\alpha  M^{(p+1)/2 + \epsilon/4}}
\end{align}
where $\alpha = \frac{\pi}{4} \sqrt{|\sin\phi_1 \cos\phi_1 |}<M$ is the constant for the given $\phi$. Substituting the $|\sin\theta_1| = {r_1}/{\sqrt{r_1^2 + H^2}}$, the range of $r_1$ satisfying \eqref{eq:range sin theta} is written as
\begin{align}\label{eq:range r}
    \sqrt{\frac{H^2}{\alpha^2 M^{2+\epsilon/2} -1}} < r_1 < 
    \sqrt{\frac{H^2}{4\alpha^2 M^{(p+1) + \epsilon/2} -1 }}.
\end{align}
Thus, the probability of $\{ Z > M^{2p}|\phi_1 \}$ is lower bounded with the probability of \eqref{eq:range r} as
\begin{align}
    &\mathbb{P}\left[Z >M^{2p} \right] \nonumber \\
    &> \mathbb{P} \left[ \sqrt{\frac{H^2}{\alpha^2 M^{2+\epsilon/2} -1}} < r_1 < 
    \sqrt{\frac{H^2}{4\alpha^2 M^{(p+1) + \epsilon/2} -1 }} \right] \label{eq:Relation Z distance} \\
    & \mathop{=}^{\text{(a)}} \exp\left[-\frac{\lambda \pi H^2}{\alpha^2 M^{2+\epsilon/2}-1} \right] - \exp \left[-\frac{\lambda \pi H^2}{4 \alpha^2 M^{(p+1) + \epsilon/2} -1 } \right] \label{eq:prob Z} \\
    & \mathop{\rightarrow}^{\text{(b)}} 1 \notag
\end{align}
where (a) is from Lemma \ref{lem:prob distance} and (b) holds for the region $q \in (p+1+\epsilon/2,2+\epsilon/2)$. 
We note that \eqref{eq:Relation Z distance} describes how the location of user affects the performance.
\\
{$\bullet$ Case 2 ($\phi = 0$)}: We obtain the case for $\vartheta^x = \sin \theta_1$ and $\vartheta^y =0$ with similar approach for $\phi_1 \neq 0$. Then, the event $\{ Z >M^{2p} | \phi_1 \}$ is given by
\begin{align} 
    \frac{1}{M^2} \left| \frac{ \sin\left(\frac{\pi M}{2} \sin\theta_1 \right) }
    { \sin\left(\frac{\pi }{2} \sin\theta_1  \right) } \right|^2 > M^{2p}.  \label{eq:M>Z chi=0}
\end{align}
We can find the sufficient conditions of \eqref{eq:M>Z chi=0} as
\begin{align}
    \left| \sin\left(\frac{\pi M}{2} \sin \theta_1 \right) \right| > \frac{1}{M^{\epsilon/4}} \notag
\end{align}
and
\begin{align}
    \left| \sin\left(\frac{\pi }{2} \sin \theta_1 \right) \right| < \frac{1}{M^{p+ \epsilon/4}}. \notag 
\end{align}
From the sufficient condition, we obtain the bound of $|\sin \theta_1|$ using the same approach when $\phi \neq 0$ as
\begin{align}\label{eq:range theta chi=0}
    \frac{1}{\frac{\pi}{4}M^{1+\epsilon/4} } < |\sin\theta_1 | < \frac{1}{\frac{\pi}{2}M^{p+\epsilon/4} }
\end{align}
with the $|\sin\theta_1| = r_1/\sqrt{r_1^2 + H^2}$, the range of $r_1$ satisfying \eqref{eq:range theta chi=0}
\begin{align}
    \sqrt{\frac{H^2}{\frac{\pi^2}{16}M^{2+\epsilon/2}-1 }} < r_1 < \sqrt{\frac{H^2}{\frac{\pi^2}{4} M^{2p+\epsilon/2}-1 }} \notag
\end{align}
Thus, the probability of the event $\{Z>M^{2p} | \phi_1 \}$ when $\phi_1 = 0$ is lower bounded as
}
}
\begin{align}
    &\mathbb{P}[Z >M^{2p}] \nonumber \\ &> \exp\left[ -\frac{\lambda \pi H^2}{\frac{\pi^2}{16} M^{2+\epsilon/2} -1 } \right] - \exp \left[ -\frac{\lambda \pi H^2}{\frac{\pi^2}{4} M^{2p+\epsilon/2} -1} \right] \notag \\
    & \mathop{\rightarrow}^{(a)} 1 \notag
\end{align}
where (a) holds when $q \in (2p+\epsilon/2, 2 + \epsilon/2)$. We note that $2p+\epsilon/2 \leq p+1+\epsilon/2$ is always satisfied with $0\leq p\leq1$.

From these results, we obtain lower and upper bounds on $\mathcal{\bar{R}}_{r}$ where $r_1$ ranges within \eqref{eq:range r} as
\begin{align}\label{eq:bound bar R}
    \text{L}_r = \mathbb{E}\left[\log \left( 1 + \frac{Z}{H^2 \left(1+\frac{1}{\alpha^2 M^{2+\epsilon/2}-1} \right)} \right) \right] < \mathcal{\bar{R}}_r  \notag \\
    < \mathbb{E}\left[\log \left( 1 + \frac{Z}{H^2 \left(1+\frac{1}{4\alpha^2 M^{(p+1)+\epsilon/2}-1} \right)} \right) \right] = \text{U}_r
\end{align}
where $\text{L}_r$ and $\text{U}_r$ are the lower and upper bound of $\mathcal{\bar{R}}_r$, respectively.
We have to note that $\mathbb{P}[ Z >M^{2p}] \rightarrow 1$ as $M \rightarrow \infty$ for $q \in (p+1+\epsilon/2,2+\epsilon/2)$. Here, by setting $p=q-1-\epsilon$, the lower bound $\text{L}_r$ is derived as
    \begin{align}
        \text{L}_r &> \int_{M^{2(q-1-\epsilon)}}^{M^2} \log\left(1+\frac{z}{H^2 \left(1+\frac{1}{\alpha^2 M^{2+\epsilon/2}-1} \right)} \right) p(z) dz \notag \\
        & \geq \log\left(1+\frac{M^{2(q-1-\epsilon)}}{H^2 \left(1+\frac{1}{\alpha^2 M^{2+\epsilon/2}-1} \right)} \right) \int_{M^{2(q-1-\epsilon)}}^{M^2} p(z) dz \notag \\
        & \rightarrow \log \left(1+\frac{1}{H^2} M^{2(q-1-\epsilon)} \right) \text{ as $M\rightarrow \infty$}. \notag
    \end{align}
    By setting $p= q-1+\epsilon$, we obtain the upper bound $\text{U}_r$ as
    \begin{align}
        \text{U}_r &= \int_{M^{2(q-1+\epsilon)}}^{M^2} \log \left( 1 + \frac{z}{H^2 \left(1+\frac{1}{4\alpha^2 M^{(p+1)+\epsilon/2}-1} \right)} \right)p(z)dz \notag \\
        &+ \int_{0}^{M^{2(q-1+\epsilon)}} \log \left( 1 + \frac{z}{H^2 \left(1+\frac{1}{4\alpha^2 M^{(p+1)+\epsilon/2}-1} \right)} \right) p(z) dz \notag \\
        &\leq \log \left( 1 + \frac{1}{H^2}M^2 \right)\int_{M^{2(q-1+\epsilon)}}^{M^2} p(z)dz \notag \\
        &\phantom{-----} +\log \left( 1 + \frac{1}{H^2}M^{2(q-1+\epsilon)} \right) \int_{0}^{M^{2(q-1+\epsilon)}}  p(z) dz \notag \\
        &\mathop{\rightarrow}^{\text{(a)}} \log\left(1+\frac{1}{H^2} M^{2(q-1+\epsilon)} \right) \text{ as $M \rightarrow \infty$} \notag
    \end{align}
    where (a) is from the fact that by using \eqref{eq:prob Z}, we know that $\mathbb{P}[ Z > M^{2p}]\rightarrow 0$ for $q < p+1-\epsilon/2$. \textcolor{black}{The condition $\epsilon = O\left(\frac{1}{\log M} \right)$ ensures the tightness of the lower and upper bounds, as it causes $\text{U}_r - \text{L}_r = 4\epsilon \log M \rightarrow 0$ as $M$ grows.}
    Then, we have
    \begin{align} 
     \log\left(1+\frac{1}{H^2} M^{2(q-1-\epsilon)} \right) < \mathcal{\bar  R}_r < \log\left(1+\frac{1}{H^2} M^{2(q-1+\epsilon)}\right). \notag
    \end{align} 
We can modify this result into 
\begin{align}
    \log\left(1+\frac{\beta}{H^2}M^{2(q-1-\epsilon)} \right) &< \mathbb{E}_{r|\phi}\left[\log\left(1 + \frac{\beta}{r_1^2+H^2} Z \right) \right] \notag \\
    &< \log\left(1+\frac{\beta}{H^2}M^{2(q-1+\epsilon)} \right) \label{eq:modify bar R}
\end{align}
where $\beta$ is independent to the distance $r_1$. From the fact $\mathcal{R}_1 = \mathbb{E}_{|g_1|^2,\phi_1}\left[ \mathbb{E}_{r_1|\phi_1}\left[\log\left(1 + P G_{\text{Tx}} G_{\text{Rx}} G_{\text{L}} \frac{|g_1|^2}{r_1^2+H^2} M^2 f_Z(r_1,\phi_1) \right) \right] \right]$ where $L_1 =  G_L \frac{1}{r_1^2+H^2}$ and $G_L = \left(\frac{c_0}{4\pi f_c } \right)^2$, the lower bound of $\mathcal{R}_1$ for $\lambda \sim M^q$ with $q \in (p+1+\epsilon/2, 2+\epsilon/2)$ for $p\in(0,1)$ is obtained by
\begin{align}
    \mathcal{R}_1  &\mathop{>}^{\text{(a)}} \mathbb{E}\left[ \log\left(1+  P G_{\text{Tx}} G_{\text{Rx}} G_{\text{L}} \frac{|g_1|^2}{H^2}  M^{2(q-1-\epsilon)} \right) \right] \notag \\
    &\mathop{=}^{\text{(b)}} \log\left( P G_{\text{Tx}} G_{\text{Rx}} G_{\text{L}} \frac{1}{H^2}  M^{2(q-1+\epsilon)} \right) + \mathbb{E}\left[ \log\left(|g_1|^2 \right) \right] \notag \\
    &\mathop{=}^{\text{(c)}} \log\left( P G_{\text{Tx}} G_{\text{Rx}} G_{\text{L}} \frac{1}{H^2}  M^{2(q-1+\epsilon)} \right) + \int_0^\infty \log x f_X(x) dx \notag \\
    &\mathop{=}^{\text{(d)}} \log M^{ 2(q-1-\epsilon)} +  \gamma \label{eq:lower bound R1}
\end{align}
where (a) holds from \eqref{eq:modify bar R} by letting $\beta = P G_{\text{Tx}} G_{\text{Rx}} G_{\text{L}} |g_1|^2$, (b) is from properties of log for large $M$, (c) is the definition of expectation where $f_X(x)$ is in \eqref{eq:SSF} and (d) holds when $\gamma = \log\left( P G_{\text{Tx}} G_{\text{Rx}} G_{\text{L}} \frac{1}{H^2} \right) + \int_0^\infty \log x f_X(x) dx$ is the constant independent to $M$.
{\color{black}{Through ergodic averaging, the randomness due to small-scale fading is marginalized into a constant term that does not scale with $M$, and is therefore effectively decoupled from the asymptotic behavior of the system. As fading is captured as a constant through expectation, the result holds under general fading assumptions.}}
The upper bound is obtained by similar approach such as
\begin{align}
    \mathcal{R}_1  &\mathop{<} \mathbb{E}\left[ \log\left(1+  P G_{\text{Tx}} G_{\text{Rx}} G_{\text{L}} \frac{|g_1|^2}{H^2}  M^{2(q-1+\epsilon)} \right) \right] \notag \\
    &\mathop{\approx} \log M^{2(q-1+\epsilon)} +  \gamma \label{eq:lower bound R1 2}
\end{align}
where $\lambda \sim M^q$ with $q \in (p+1+\epsilon/2, 2+\epsilon/2)$ for $p\in(0,1)$. This proof refers to \cite{gilwon:twc:2016}.
Then, the derived lower and upper bound of $\mathcal{R}_1$ is given by
\begin{align} \label{eq:bound R1}
    \log M^{2(q-1-\epsilon)} +  \gamma < \mathcal{R}_1 < \log M^{2(q-1+\epsilon)} +  \gamma.
\end{align}
\textcolor{black}{By focusing solely on the parameters of our interest and neglecting the constant \( \gamma \), we complete the proof.}

\section{Proof of Theorem \ref{thm:single fraction}} \label{appendix:proof coro single}
The denominator $\mathbb{E} \left[\log\left(1 +  P G_{\text{Tx}} G_{\text{Rx}} G_{\text{L}}   \frac{|g_1|^2}{r_1^2 + H^2} M^2 \right) \right]$ where $L_1 =  G_L \frac{1}{r_1^2+H^2}$ and $G_L = \left(\frac{c_0}{4\pi f_c } \right)^2$ is obtained by
\begin{align}
    & \mathbb{E} \left[\log\left(1 +  P G_{\text{Tx}} G_{\text{Rx}} G_{\text{L}}   \frac{|g_1|^2}{r_1^2 + H^2} M^2 \right) \right] \notag \\
    &\mathop{\approx }^{(\text{a})}   \mathbb{E}\left[ \log \left( P G_{\text{Tx}} G_{\text{Rx}} G_{\text{L}} \frac{1}{r^2+H^2} M^2 \right) \right] + \mathbb{E}\left[\log |g|^2 \right] \notag \\
    &\mathop{=}^{\text{(b)}} \mathbb{E}\left[ \log \left( P G_{\text{Tx}} G_{\text{Rx}} G_{\text{L}} \frac{1}{H^2} M^2 \right) \right] + \int_0^\infty \log x f_X(x) dx \notag \\
    &\mathop{=} \log M^2 + \gamma \label{eq: max R1}
\end{align}
where (a) holds for large $M$. 
The first term of (b) comes from Appendix \ref{appendix:proof single r}, the second term is in \eqref{eq:lower bound R1}. We divide the result of Theorem \ref{thm single r} by $\mathbb{E} \left[\log\left(1 +  P G_{\text{Tx}} G_{\text{Rx}} G_{\text{L}}   \frac{|g|^2}{r^2 + H^2} M^2 \right) \right]$ using the result \eqref{eq: max R1} such as
\begin{align}
    \frac{2(q-1-\epsilon)\log M +  \gamma}{2\log M + \gamma} 
    &< \frac{\mathcal{R}_1}{\mathbb{E} \left[\log\left(1 +  P G_{\text{Tx}} G_{\text{Rx}} G_{\text{L}}  \frac{|g|^2}{r^2 + H^2} M^2 \right) \right]} \notag \\
    &<  \frac{2(q-1+\epsilon)\log M +  \gamma}{2\log M + \gamma}. \notag
\end{align}
As $M \rightarrow \infty$, we have
\begin{align}
     q-1-\epsilon &< \frac{\mathcal{R}_1}{\mathbb{E} \left[\log\left(1 +  P G_{\text{Tx}} G_{\text{Rx}} G_{\text{L}}   \frac{|g|^2}{r^2 + H^2} M^2 \right) \right]} < q-1+\epsilon
\end{align}
with small positive $\epsilon$, we conclude the proof.

\section{Proof of Corollary \ref{coro:MU single beam}} \label{appendix:MU single beam}
Assuming a random variable $X=|g_1|^2$ where $g \sim \CMcal{SR}(\Omega, b_0, m)$, we have
    \begin{align}
        & \mathbb{E}\left[ \log \left(1 + \frac{X h_{1}(r_1,\phi_1) }{\sum_{i \neq 1} X h_{i}(r_1,\phi_1) + 1} \right) \right] \notag \\
        & \mathop{=} \mathbb{E}\left[ \mathbb{E}\left[\log \left(1 + \frac{X h_{1}(r_1,\phi_1) }{\sum_{i \neq 1} X h_{i}(r_1,\phi_1) + 1} \right) \middle| r_1,\phi_1 \right] \right] \notag \\
        & \mathop{=}^{(\text{a})} \int_0^R \int_0^{2\pi} \int_0^\infty \frac{1}{\tau} \left( \mathbb{E}\left[ e^{-\tau X \sum_{i\neq 1} h_{i}(r,\phi)} \right] \right.   \notag \\
        & \phantom{------} \left. -\mathbb{E} \left[ e^{-\tau X \sum_{i=1} h_{i}(r,\phi)}\right] \right) e^{-\tau} \lambda r e^{-\pi \lambda r^2} d\tau d\phi dr \notag \\
        &\mathop{=}^{(\textcolor{black}{\text{b}})}  \int_0^R \int_0^{2\pi} \int_0^\infty \frac{1}{\tau} \left( \frac{(2b_0m)^m (1+2 \tau b_0 \sum_{i\neq 1} h_{i}(r,\phi) )^{m-1}}{[(2b_0m + \Omega)(1+2 \tau b_0 \sum_{i\neq 1} h_{i}(r,\phi) )-\Omega]^m } \right. \nonumber  \\ 
        &  \left. - \frac{(2b_0m)^m (1+2 \tau b_0 \sum_{i = 1} h_{i}(r,\phi) )^{m-1}}{[(2b_0m + \Omega)(1+2 \tau b_0 \sum_{i = 1} h_{i}(r,\phi) )-\Omega]^m } \right) \nonumber \\
        &e^{-\tau} \lambda r e^{-\pi \lambda r^2} d\tau d\phi dr \notag    
        \end{align}
    where (a) comes from a useful lemma in \cite{hamdi:tcom:10}
    \begin{align}
        &\mathbb{E}\left[\log \left(1+\frac{x_k}{\sum_{i\neq 1}x_i + 1} \right)\right] \notag \\
        &= \int_0^\infty \frac{1}{\tau} \left(\mathbb{E}\left[e^{-\tau \sum_{i\neq 1}x_i}\right]-\mathbb{E}\left[e^{-\tau \sum_{i=1}x_i}\right] \right)e^{-\tau}d\tau
    \end{align}
    and (b) is from the definition of MGF given in \eqref{eq:SR mgf}. By substituting $h_{i}(r,\phi) = \bar{P} G_{\text{Tx}} G_{\text{Rx}} L_1 M^2 f_{Z_{i}}(r,\phi)$, we conclude the proof.

\section{Proof of Lemma \ref{lem:single interf}} \label{proof of lem:single interf}
For $s,\ell \in (0,1)$, the event $\left\{ \frac{M^2 }{M^{2\ell}} f_{Z_i}(r_1,\phi_1) < \frac{1}{M^{2s}} \right\}$ is equivalent to
\begin{align}
     M^{2(1-\ell)}  F^2_M(\vartheta^x_{i} - \vartheta^x_{1}) F^2_M(\vartheta^y_{i} - \vartheta^y_{1}) < \frac{1}{M^{2s}} \label{appendix:eq:proof interf 1} 
\end{align}
where $ f_{Z_i}(r_1,\phi_1) = F^2_M(\vartheta^x_{i} - \vartheta^x_{1}) F^2_M(\vartheta^y_{i} - \vartheta^y_{1})$.
We have the two sufficient conditions for \eqref{appendix:eq:proof interf 1} as
\begin{align}
    M^{1-\ell} F_M^2(\vartheta_i^x - \vartheta_1^x) < \frac{1}{M^s} \label{eq:interf suff1}
\end{align}
and 
\begin{align}
    M^{1-\ell} F_M^2(\vartheta_i^y - \vartheta_1^y) < \frac{1}{M^s} \label{eq:interf suff2}
\end{align}
Here, focusing on \eqref{eq:interf suff1}, we have
\begin{align}
    \frac{1}{M^{1+\ell}} \left| \frac{ \sin \frac{\pi M}{2} (\vartheta_{i}^x -\vartheta_1^x )  }{ \sin \frac{\pi}{2} (\vartheta_{i}^x -\vartheta_1^x ) } \right|^2 < \frac{1}{M^s}.
\end{align}
By using $|\sin \frac{\pi M}{2} (\vartheta_{i}^x -\vartheta_1^x )| < 1$, we have
\begin{align}
    \left|\sin \frac{\pi}{2} (\vartheta_{i}^x -\vartheta_1^x )\right|> \frac{1}{M^{(1+\ell-s)/2}}.
\end{align}
From the fact that $|\sin x | > \frac{x}{2}$ for $|x| \in (0,\pi/2)$, we have
\begin{align}
    \frac{\pi}{4} \left| \vartheta_{i}^x -\vartheta_1^x  \right| >  \frac{1}{M^{(1+\ell-s)/2}}.
\end{align}
and with $\vartheta_{i}^x = \frac{2n}{M^\ell}$ and $\vartheta_1^x = \sin \theta_1 \cos \phi_1$ 
\begin{align}
    \left| \frac{2n}{M^\ell} - \sin \theta_1 \cos \phi_1  \right| > \frac{1}{\frac{\pi}{4} M^{(1+\ell-s)/2} }. \label{eq:interf rearrange result}
\end{align}
Relaxing in terms of absolute value, we have
\begin{align}
    \left| \frac{2n}{M^\ell} - \sin \theta_1 \cos \phi_1  \right| \mathop{\geq}^{\text{(a)}} \left| \frac{2n}{M^\ell} \right| - \left|\sin \theta_1 \cos \phi_1  \right| \mathop{\geq}^{\text{(b)}} \left| \frac{2n}{M^\ell} \right| - \left|\sin \theta_1  \right| \notag
\end{align}
where $\left|\sin \theta_1 \cos \phi_1  \right|$ are $[u,v]$ coordinates of user $1$ in the nadir-pointing beam's coverage and $\left| \frac{2n}{M^\ell} \right|$ is always located at the point outside the nadir-pointing beam's coverage. (b) comes from $|\cos\phi_1|\le 1$. Then, we have
\begin{align}
    |\sin\theta_1| < \left| \frac{2n}{M^\ell} \right| - \frac{1}{\frac{\pi}{4} M^{(1+\ell-s)/2} } 
\end{align}
and for $\ell+s < 1$ with sufficiently large $M$
\begin{align}\label{eq:suff result of I}
     |\sin\theta_1| < \frac{1}{M^\ell} 
\end{align}
The procedure of sufficient condition \eqref{eq:interf suff2} is almost similar with \eqref{eq:interf suff1} and given by
\begin{align}
    \left| \frac{2m}{M^\ell} - \sin \theta_1 \sin \phi_1  \right| > \frac{1}{\frac{\pi}{4} M^{(1+\ell-s)/2} } \label{eq:interf rearrange result2}
\end{align}
which is corresponding to \eqref{eq:interf rearrange result}. And, the result is equal to \eqref{eq:suff result of I}.
Therefore, the probability of \eqref{appendix:eq:proof interf 1} is given by
\begin{align}
    \mathbb{P}\left[ \textcolor{black}{\frac{M^2}{M^{2\ell}}  f_{Z_i}(r_1,\phi_1)} < \frac{1}{M^{2s}}\right] &> \mathbb{P}\left[ |\sin\theta_1| <  \frac{1}{M^\ell}    \right] \notag \\
    &\mathop{=}^{(\text{a})}  \mathbb{P}\left[ r_1 <  \frac{H}{\sqrt{M^{2\ell}-1}}    \right] \notag \\
    &\mathop{=}^{(\text{b})}  1 - \exp\left[ - \frac{\lambda \pi H^2}{M^{2\ell}-1} \right]
\end{align}
where (a) is $\sin \theta_1 = r_1/\sqrt{r_1^2 + H^2}$ and (b) is using Lemma \ref{lem:prob distance}. This is the end of proof.

\section{Proof of Theorem \ref{thm:single interf rate}} \label{appendix:single interf rate}
   The lower bound of $\mathcal{R}^M_1$ is given by
\begin{align}
    &\mathbb{E}\left[ \log \left( 1 + \frac{ \bar{P} G_{\text{Tx}} G_{\text{Rx}} L_1 |g_1|^2 M^2 f_{Z_1}(r_1,\phi_1)  }{ \bar{P} G_{\text{Tx}} G_{\text{Rx}} L_1 |g_1|^2 \sum_{i\neq 1} M^2 f_{Z_{i}}(r_1,\phi_1)  + 1} \right) \right] \notag \\
    &\mathop{\approx}^{(\text{a})} \mathbb{E}\left[ \log \left( \frac{ {P} G_{\text{Tx}} G_{\text{Rx}} L_1 |g_1|^2 M^{2-2\ell} f_{Z_{1}}(r_1,\phi_1)  }{ {P} G_{\text{Tx}} G_{\text{Rx}} L_1 |g_1|^2 \sum_{i\neq 1} M^{2-2\ell} f_{Z_{i}}(r_1,\phi_1)  + 1} \right) \right] \notag \\
    &\mathop{>}^{(\text{b})} \mathbb{E}\left[ \log \left(  \frac{ {P} G_{\text{Tx}} G_{\text{Rx}} L_1 |g_1|^2 M^{2-2\ell} f_{Z_{1}}(r_1,\phi_1)  }{ {P} G_{\text{Tx}} G_{\text{Rx}} L_1 |g_1|^2 \sum_{i\neq 1} \mathbb{E}\left[M^{2-2\ell} f_{Z_{i}}(r_1,\phi_1) \right]  + 1} \right) \right] \notag \\
    &\mathop{>}^{(\text{c})} \mathbb{E}\left[ \log \left(  \frac{ {P} G_{\text{Tx}} G_{\text{Rx}} L_1 |g_1|^2 M^{2-2\ell} f_{Z_{1}}(r_1,\phi_1)  }{ {P} G_{\text{Tx}} G_{\text{Rx}} L_1 |g_1|^2 \sum_{i\neq 1} \frac{1}{M^{2s}}  + 1} \right) \right] \notag \\
    &\mathop{>}^{(\text{d})} \mathbb{E}\left[ \log \left(  {P} G_{\text{Tx}} G_{\text{Rx}} L_1 |g_1|^2 M^{2-2\ell} f_{Z_{1}}(r_1,\phi_1) \right) \right] \notag \\
    &\mathop{>}^{(\text{e})} \log M^{2(q-\ell - 1-\epsilon)} +  \gamma \label{eq:lower bound interf rate}
\end{align}
where (a) comes from $\text{SINR}_1 \gg 1 $ with $\bar{P} = \frac{P}{M^{2\ell}}$. (b) holds by the Jensen's inequality, (c) is Lemma \ref{lem:single interf} for $s \in (0,1)$ and $\ell \in (0, 1)$ such that $\ell + s <1$, (d) holds for sufficiently large $M\gg {P} G_{\text{Tx}} G_{\text{Rx}} L_1|g_1|^2 $, and (e) is straightforward referring to \eqref{eq:lower bound R1}. The upper bound is obtained with similar approach as
\begin{align}
    &\mathbb{E}\left[ \log \left( 1 + \frac{ \bar{P} G_{\text{Tx}} G_{\text{Rx}} L_1 |g_1|^2 M^2 f_{Z_{1}}(r_1,\phi_1)  }{ \bar{P} G_{\text{Tx}} G_{\text{Rx}} L_1 |g_1|^2 \sum_{i\neq 1} M^2 f_{Z_{i}}(r_1,\phi_1)  + 1} \right) \right] \notag \\
    &\mathop{\approx}^{(\text{a})} \mathbb{E}\left[ \log \left( \frac{ {P} G_{\text{Tx}} G_{\text{Rx}} L_1 |g_1|^2 M^{2-2\ell} f_{Z_{1}}(r_1,\phi_1)  }{ {P} G_{\text{Tx}} G_{\text{Rx}} L_1 |g_1|^2 \sum_{i\neq 1} M^{2-2\ell} f_{Z_{i}}(r_1,\phi_1)  + 1} \right) \right] \notag \\
    &\mathop{<}^{(\text{b})} \mathbb{E}\left[ \log \left( \bar{P} G_{\text{Tx}} G_{\text{Rx}} L_1 |g_1|^2 M^{2-2\ell} f_{Z_{1}}(r_1,\phi_1) \right) \right] \notag \\
    &\mathop{<}^{(\text{c})} \log M^{2(q-\ell - 1 + \epsilon)} +  \gamma
\end{align}
where (a) comes from $\text{SINR}_1 \gg 1 $ with $\bar{P} = \frac{P}{M^{2\ell}}$. (b) holds by ignoring the interference, (c) is also from \eqref{eq:lower bound R1 2}. The derivation is straightforward referring to \eqref{eq:lower bound interf rate}. Then, the bound of $\mathcal{R}^M_1$ is given by
\begin{align}
    \log M^{2(q-\ell - 1-\epsilon)} +  \gamma < \mathcal{R}^M_1 < \log M^{2(q-\ell - 1 + \epsilon)} +  \gamma
\end{align}
where $\gamma$ is the constant independent to $M$. For $M\rightarrow \infty$ with small positive $\epsilon$, we conclude the proof.
\section{Proof of Theorem \ref{thm:sum rate}} \label{appendix:coro sum rate}
From \eqref{eq:sum rate}, we easily obtain the lower bound of $\mathcal{R}_\Sigma$ such as $\mathcal{R}_\Sigma \ge K \mathcal{R}^M_1$. Then, we have 
\begin{align}
    K\log M^{2(q-\ell-1-\epsilon)} < K \mathcal{R}^M_1 < K \log M^{2(q-\ell-1+\epsilon)}. \label{appendix:coro2:eq:K R_1}
\end{align}
By substituting $P_0 = P_0/K$ in \eqref{eq: max R1}, we can easily derive the following as
\begin{align}
     \mathbb{E}\left[ \log \left(1 + \bar{P} G_{\text{Tx}} G_{\text{Rx}} L_k |g_k|^2 M^2  \right) \right]  = \log M^{2(1-\ell)} + \gamma,
  \label{appendix:coro:eq:single}
\end{align}
where $\gamma$ is the constant independent to $M$ defined in \eqref{eq:lower bound R1}. 
By dividing \eqref{appendix:coro2:eq:K R_1} by \eqref{appendix:coro:eq:single}, for $M\rightarrow \infty$ we have
\begin{align}
    \frac{\log M^{2(q-\ell-1-\epsilon)}}{\log M^{2(1-\ell)} } &< \frac{K \mathcal{R}^M_1}{K\cdot\mathbb{E}\left[ \log \left(1 + \bar{P} G_{\text{Tx}} G_{\text{Rx}} L_1 |g_1|^2 M^2  \right) \right]}  \notag \\
    &\phantom{---------} < \frac{\log M^{2(q-\ell-1+\epsilon)}}{\log M^{2(1-\ell)} } 
\end{align}
with sufficiently small $\epsilon$, we conclude the proof.
\bibliographystyle{IEEEtran}
\bibliography{main}

\end{document}